\documentclass[reprint,amsmath,amssymb,aps,superscriptaddress]{revtex4-1}
\usepackage{graphicx}
\usepackage{color,soul}
\usepackage[caption=false]{subfig}
\usepackage{amsmath}
\usepackage{float}
\usepackage[normalem]{ulem}
\usepackage{xr-hyper}
\usepackage{romannum}
\usepackage{gensymb}
\usepackage{makecell}
\usepackage{setspace}
\usepackage{placeins}

\usepackage[colorlinks=true,linkcolor=blue,anchorcolor=black,
citecolor=magenta,filecolor=blue,menucolor=black,
runcolor=black,urlcolor=blue]{hyperref}

\newcommand{\SK}[1]{\textcolor{black}{{#1}}}

\begin{document}
\pagenumbering{arabic}
\title{Enhanced Long Wavelength Mermin-Wagner-Hohenberg Fluctuations in Active Crystals and Glasses}
\author{Subhodeep Dey}
\thanks{equal contributions}
\affiliation{Tata Institute of Fundamental Research Hyderabad, 36/P, Gopanpally Village, Serilingampally Mandal, Ranga Reddy District,
Hyderabad, Telangana 500046, India}
\author{Antik Bhattacharya}
\thanks{equal contributions}
\affiliation{Tata Institute of Fundamental Research Hyderabad, 36/P, Gopanpally Village, Serilingampally Mandal, Ranga Reddy District,
Hyderabad, Telangana 500046, India}
\author{Smarajit Karmakar}\email{smarajit@tifrh.res.in}
\affiliation{Tata Institute of Fundamental Research Hyderabad, 36/P, Gopanpally Village, Serilingampally Mandal, Ranga Reddy District,
Hyderabad, Telangana 500046, India}
\begin{abstract}
\SK{In two-dimensions (2D), the Mermin-Wagner-Hohenberg (MWH) fluctuation plays a significant role, giving rise to striking dimensionality effects marked by long-range density fluctuations leading to the singularities of various dynamical properties.  According to the MWH theorem, a 2D equilibrium system with continuous degrees of freedom cannot achieve long-range crystalline order at non-zero temperatures. Recently, MWH fluctuations have been observed in glass-forming liquids, evidenced by the logarithmic divergence in the plateau value of mean squared displacement (MSD). Our research investigates long-wavelength fluctuations in crystalline and glassy systems influenced by non-equilibrium active noises. Active systems serve as a minimal model for understanding diverse non-equilibrium dynamics, such as those in biological systems and self-propelled colloids. We demonstrate that fluctuations from active forces can strongly couple with long-wavelength density fluctuations, altering the lower critical dimension ($d_l$) from $2$ to $3$ and leading to a novel logarithmic divergence of the MSD plateau with system size in 3D.}
\end{abstract} 
\maketitle
\noindent{\bf\large Introduction: }The Mermin-Wagner-Hohenberg (MWH) theorem \cite{Hohenberg1967, Mermin1966, Mermin1968} asserts that continuous spontaneous symmetry breaking (SSB) cannot occur at any finite temperature $(T>0)$ in a 2D equilibrium system. Hohenberg first introduced this for superfluid systems, and later, building on his work, Mermin and Wagner demonstrated that phase transitions involving continuous symmetry breaking are likewise impossible in 2D equilibrium systems consist of continuous spin or particles. In particular, there is no long-range order in the system due to the instability caused by the long wavelength fluctuations in the system below the lower critical dimension ($d_l$); these long wavelength fluctuations are often termed as Mermin-Wagner-Hohenberg (MWH) fluctuations in the literature.  As MWH theory concerns only the physics of large length scale or small wave vector, it is natural to expect that the theory will be true for both crystalline and disordered solids as long as long wavelength phononic excitations are present in the system. It is often argued in the theoretical derivation of the Mermin-Wagner-Hohenberg (MWH) theorem \cite{Hohenberg1967, Mermin1966, Mermin1968} that the dispersion behaviour of phonons determines the thermal vibrations of the molecules or particles in their equilibrium crystalline position \cite{Jancovici1967, Imry1971}, and it is quite straightforward to show that the mean squared displacement (MSD) of particles from their equilibrium position, the Debye-Waller (DW) Factor, diverges logarithmically with the linear dimension ($L$) of the system in 2D.  Recently, the MWH fluctuations are found to be dominant in both supercooled liquids and amorphous solids, Shiba et al.  first showed this diverging fluctuation effect for 2D disordered system using computer simulation \cite{Shiba2016}. For 2D colloidal systems, the presence of such fluctuations has been confirmed by recent works \cite{Flenner2015, Vivek2017, Illing2017}. Again,  Li et al. \cite{Li2019} have shown that long wavelength density fluctuations also influence the dynamics in the liquid states in these 2D systems.

In \cite{Li2019} it was shown that MWH fluctuations can have significantly strong effects even at high temperatures manifested by the breakdown of the Stokes-Einstein (SE) relation in a manner that is different from the breakdown observed in the supercooled temperature regimes. SE relates the diffusivity ($D$) of the particles in the medium with the characteristic relaxation time ($\tau_\alpha$) or viscosity ($\eta$) as $D = K_B T/C \eta$, with $K_B$ being the Boltzmann constant, and $C$ being a constant that depends on the details of the probe particle. Assuming $ \tau_\alpha \simeq \eta/K_B T$ \cite{shi2013relaxation}, one gets $D\sim 1/\tau_\alpha$.  On the other hand,  in the supercooled temperature regime, one finds: $D \propto \tau_\alpha ^{-\kappa}$, with $\kappa$ being smaller than $1$.  This is often referred to as fractional SE relation. Fractional SE is known from experiments in various molecular glass-forming liquids and model glass-forming systems in simulations. SE breakdown in 3D can be often explained using the concept of growing dynamic heterogeneity (DH) in these systems ~\cite{Paul2023}.  Whereas in 2D, the picture is very different even at high temperatures with $\kappa > 1$ is reported in the high-temperature normal liquid regimes and $\kappa < 1$ being reported in the supercooled temperature regime similar to the 3D case. The observation of $\kappa > 1$ at high-temperature liquids in 2D has been shown as purely coming from the MWH like long-wavelength density fluctuations in the system, thereby proving that MWH fluctuations are prevalent even at high temperatures as much as in supercooled and low temperature solid regimes \cite{Li2019}. \SK{This suggests that study of SE breakdown at high temperatures can be a good way to probe the exisitence of MWH-like fluctuations in experiments.}

A hallmark characteristic of glassy dynamics is the enhanced dynamical heterogeneity, which quantifies the marked difference in relaxation patterns in different parts of a glassy system \cite{Cavagna2009, Berthier2011, Karmakar2014}. Dynamic heterogeneity is often highlighted as one of the important features of glassy dynamics. Recent research \cite{Karmakar2016} has shown that the DH of glassy systems, calculated using dynamic susceptibility ($\chi_4(t)$; see the Methods Section for its definition), exhibits additional short-time peaks in three-dimensional (3D) systems. These peaks are attributed to collective motions facilitated by long-wavelength phonon modes \cite{Karmakar2016}.  The short time peak of $\chi_4(t)$ disappears if one does Brownian dynamics simulation of the same systems in the over-damped limit where the phonons get suppressed considerably. Notably, these short-time peaks are found to be further enhanced in the presence of colored active noise at 3D, suggesting that active noise play an important role in amplifying phonon modes even in 3D \cite{Dey2022}. Thus, it is natural to expect that 2D systems will be influenced even more strongly by these long-wavelength phonon modes when active noises are present. Many experiments are conducted in 2D or quasi-2D geometries due to practical experimental considerations. Even biological systems, such as cell monolayers, which show glass-like dynamics, can be considered quasi-2D systems. Therefore, understanding the presence of MWH fluctuations in non-equilibrium systems like active glasses or active crystalline solids in various dynamical conditions (damping),  becomes crucial as synthetic active colloidal particles are in a dynamical regime where inertia is important whereas cell monolayers are in the overdamped regime where inertia can be neglected. Significance of inertial effect in active system has been shown in different works \cite{Cavagna2014,Cavagna2018,Dey2022,teVrugt2023}.  Furthermore, investigating whether the presence of these fluctuations leads to increased or suppressed mean squared position fluctuations in the system is essential for interpreting the dynamical information obtained from these 2D or quasi-2D systems.

In recent times, there has been a surge of research activities in the field of active matter, leading to the emergence of a new direction of studies on disordered systems called active glasses \cite{Mandal2016, Paul2023}. Active matter is often categorized as a system in which the constituents can move internally, driven by their internal energy, in addition to the environmental influence of thermal fluctuations ~\cite{Vicsek1995, Toner1998, Ramaswamy2010, Palacci2013, Marchetti2013}. Such systems exhibit a plethora of interesting dynamical phenomena, including spontaneous symmetry breaking of the rotational order in two dimensions \cite{Vicsek1995} leading to the formation of ordered phases of clusters or flocks. These clusters have coherent collective motion at low noise strength and high particle density \cite{Toner1998}. Many biological systems exhibit collective dynamical behavior, in which forces generated by ATP consumption drive the dynamics instead of thermal fluctuations. A simple model of these systems that can capture some of the salient dynamical behaviors is a collection of self-propelled particles (SPPs) \cite{Marchetti2013}. Several studies show that collective dynamics of cells and tissues during cell proliferation, cancerous cell progression, and wound healing \cite{Zhou2009, Angelini2011, Parry2014, Park2015, Garcia2015, Malinverno2017, Nishizawa2017, Kim2020, Vishwakarma2020, Cerbino2021} have dynamical features similar to glassy dynamics. Cell cytoplasm and bacterial cytoplasm show glassy dynamics, which can also modulate the depletion of ATP \cite{Zhou2009, Parry2014}. These intricate dynamical similarities between various biological systems and glassy systems have fuelled a lot of research activities in modelling active glassy systems to develop an understanding of the emergent dynamical behaviors that are not very sensitive to the details of the system. Instead, they are outcomes of intrinsic non-equilibrium driving due to active forces \cite{Loi2008, Fodor2016}.

Active systems are inherently out of equilibrium in nature as they do not follow detailed balance and are driven either internally or by external forcing. In recent years, there have been attempts to understand their steady-state dynamical behavior within equilibrium statistical mechanics using an appropriate effective temperature and generalized fluctuation-dissipation theory (FDT). The work \cite{Loi2008} shows that in the small activity limit, one can define the effective temperature of the out-of-equilibrium system using the effective FDT analysis in which the time-reversal symmetry is still intact \cite{Fodor2016}. Analytical results on the dynamics of an active particle in a harmonic potential, the active Ornstein-Ulhenbeck process, suggest that in a short persistent time limit, an effective potential of the active system can be used to define effective Boltzmann weight. Similar ideas have been extended to active glasses, which suggest that some dynamical aspects can be well understood using an effective temperature description \cite{Nandi2018, Berthier2013, Berthier2014, Ni2013, Mandal2016, Nandi2017, Berthier2019}. However, higher-order dynamical correlation functions like four-point susceptibility ($\chi_4(t)$) cannot be understood with the same framework, as reported in \cite{Paul2023}. Thus, a clear understanding of active systems in their dynamical steady states is still lacking, and the intricate effects of active driving on the dynamics continue to puzzle the scientific community.

\begin{figure*}[!htb] 
	\includegraphics[width=1.0\linewidth]{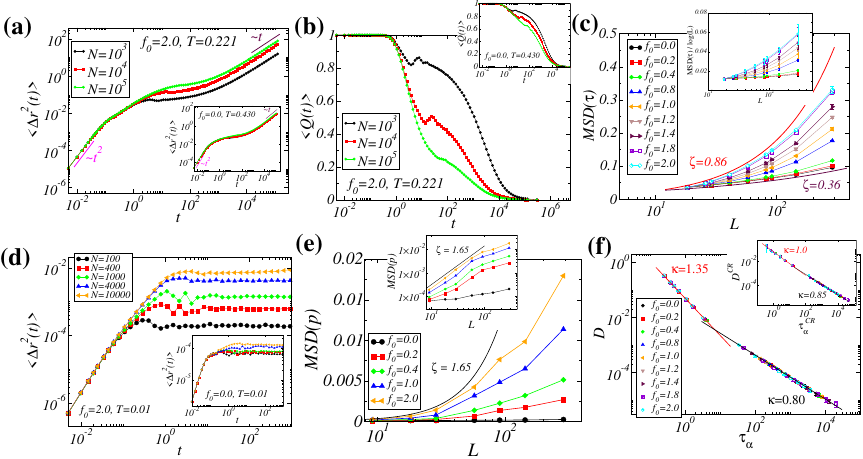}
	\caption{{\bf Mean Square Displacement and Mermin-Wagner-Hohenberg Fluctuations.} (a) Mean square displacement (MSD) as a function of time for activity $f_0=2.0$ at temperature T=0.221 with system size ranging from $10^3$ to $10^5$, here it shows increase in MSD plateau much faster than the passive system (Inset) at temperature T=0.430. (b) Two point density correlation Q(t) shows faster relaxation for the case of active system than the passive system (Inset) for large system size. (c) MSD($\tau$) plateau diverges faster than log(L) increasing activity. Inset: At larger activity it starts to power-law behavior unlike log(L) behavior of its passive counter-part. For $f_0=2.0$ the power exponent is $\zeta=0.86$. (a), (b) \& (c) are for 2dmKA system. (d) MSD as function of time for active polycrystalline system for $f_0=2.0$ at temperature T=0.01 for system ranging from $10^2$ to $10^4$, it shows faster increase in MSD plateau than passive system (Inset) at same temperature similar to (a). (e) For active polycrystal we can get similar power-law of MSD plateau divergence, here for activity $f_0=2.0$ the exponent is $\zeta=1.65$. (f) Diffusivity as a function of relaxation time shows a power-law exponent $\kappa=1.35$. this breakdown of Stokes-Einstein relation ($\kappa>1.0$) is possible due to the presence of long wavelength phonon fluctuation in 2D, which is also valid for active system as well. We again get back the Stokes-Einstein relation with $\kappa\simeq1.0$ when we do cage-relative diffusivity and relaxation time calculations, again showing the presence of phonon like excitations in active liquids as well. Error bars in the figure panels are measured by computing the standard deviation (SD) of fluctuations in various statistically independent simulations.}
	\label{fig:MerminWagner}
\end{figure*}

In this article, we have done extensive studies to understand the effect of activity on the dynamics of a model polycrystalline solid and two glass-forming model systems in 2D and 3D. We have also done simulations of a model glassy system in four spatial dimensions (4D) to establish the shift of $d_l$ to higher dimensions. These glassy models are referred to as modified Kob-Anderson model in 2D (2dmKA) with binary composition of $65:35$ and Kob-Andersen model in 3D (3dKA) with the composition ratio $80:20$ \cite{Das2017} (see Methods section). For the polycrystalline solid, we have taken a mono-atomic system and generated the solid by cooling the liquid from high temperature melt. Our simulations are carried out in canonical ensemble, with system sizes ranging from $N = 100$ to $10^5$ particles. To introduce activity into the system, we use run and tumble particle (RTP) dynamics \cite{Mandal2016, Kallol2021, Dey2022}, which can be tuned using three parameters: $c$, $f_0$, and $\tau_p$. The concentration of active particles is $c$, which is the ratio of the number of active particles ($N_a$) with respect to the total number of particles ($N$) in the system, i.e., $c=N_a/N$. $c$, is varied in the range $c \in [0, 0.6]$, while the strength of the active force applied to each particle, $f_0$, is varied in the range $f_0 \in [0, 2.5]$. The persistent timescale, $\tau_p$, determines the duration for which the active forces act along a fixed but random direction, and is varied in the range $\tau_p \in [0, 100]$ in our simulations. To understand general applicability of our results across model systems, we have also performed simulations with active Brownian particles (ABPs) in both damped and overdamped dynamical conditions. Note these systems are also controlled by the same three activity parameters namely, $f_0$, $c$ and $\tau_p$, only the details of how active forces modify the equations of motion are different, as discussed later. Further details of the models and simulation protocols are provided in the Method section.

To highlight our major findings, we showed that the MWH fluctuation is at play even at higher temperatures, consistent with the findings in passive systems. We have also demonstrated the impact of long wavelength excitations in these systems by calculating an effective dynamical matrix and showing that the \SK{the results in non-equilibrium can be understood using an effective medium theory using the same MWH arguments}, but with enhanced effect. We found that the Debye-Waller (DW) factor diverges as a power-law with increasing activity, instead of the usual logarithmic divergence seen in equilibrium 2D systems. Interestingly, in 3D we discovered novel logarithmic divergence of DW factor for non-equilibrium systems, in contrast to the absence of any divergence in equilibrium systems. We have demonstrated the robustness of our results across various active matter models at different dynamical conditions. We have also found that the effective phonon dispersion relation becomes non-linear with respect to the wave vector as activity increases, in both the dimensions leading to the shift of lower critical dimensions from $d_l = 2$ to $3$.

\vskip +0.1in
\noindent{\bf \large Results: }To explore the enhanced effect of long wavelength phonon modes in active crystals and glasses, we first computed the mean squared displacement (MSD), $\langle\Delta r^2(t)\rangle$ (see Methods for the definition), as a function of time. For 2dmKA model glass-forming liquids with activity strength $f_0 = 2.0$ at a reduced temperature of $T=0.221$, we show the results of increasing MSD plateau with system sizes $N = 10^3, 10^4$, and $10^5$ in Fig.\ref{fig:MerminWagner}(a). The inset of the same panel shows the data for passive glass-forming liquid at $T=0.450$. We observed a significant increase in the plateau value of MSD for active systems compared to the passive system at similar relaxation times, demonstrating the enhanced effect of long wavelength fluctuations due to active forcing. In Fig.\ref{fig:MerminWagner}(b), we explored the decay profile of the two-point density correlation function $Q(t)$ (see Methods for the definition) for the same system sizes as in panel A. This was done to demonstrate how long wavelength fluctuations lead to faster relaxation in active systems for larger system sizes. The inset shows the relaxation profile for passive systems. Next, we have computed the MSD plateau of the active glass-forming liquids in the supercooled regime with changing system size for different activity $f_0$, while keeping the concentration of active particles ($c = 0.1$) and persistent time ($\tau_p = 1.0$) fixed. The results are presented in Fig.\ref{fig:MerminWagner}(c). We defined the plateau value of MSD as MSD($\tau$) with $\tau$ being the time at which MSD shows a point of inflection in the plateau regime. The Debye-Waller (DW) factor is proportional to the MSD plateau MSD($\tau$). We observed logarithmic divergence of the MSD($\tau$) with system size, i.e., $MSD(\tau) \sim \log(L)$ for passive system and in the presence of activity the divergence of MSD($\tau$) increases faster than logarithmic and follows power law divergence for higher activity, i.e., $MSD(\tau) \sim L^{\zeta}$. The exponent $\zeta$ seems to be increasing systematically with increasing activity $f_0$ reaching to $\zeta \sim 0.9$ for $f_0 = 2.0$. We obtained similar results by varying $c$ \SK{and $\tau_p$} while keeping $f_0$ constant \SK{(see Supplementary Note XV for additional data)}. Thus, the results appear to be independent of the particular choice of the activity parameter.

These strong effect of long wavelength fluctuations in disordered systems, led us to investigate their effect on active polycrystalline solids. Our findings, presented in Fig. \ref{fig:MerminWagner}(d), show that the plateau value of the mean squared displacement (MSD) significantly increases with system size in the presence of activity ($f_0 = 2.0$), while for passive systems, the growth of MSD plateau follows a logarithmic divergence (see inset). For the polycrystalline model, we keep temperature of the system at $T = 0.01$ to ensure that the solid remains in the polycrystalline minimum. We also plotted the MSD plateau values as a function of $L$ in a double logarithmic plot, presented in panel E of the same figure, which shows that the divergence of MSD plateau or the Debye-Waller factor in active crystals grows as power-law in system size, similar to the results found in disordered systems. The exponent $\zeta \sim 1.65$ seems to be stronger than the exponent obtained in disordered solids. These results suggest that even for out-of-equilibrium system the activity induced enhancement of the MWH fluctuations is not sensitive to details of the structural ordering.

\begin{figure}[!htpb] 
\centering
	\includegraphics[scale = 0.61]{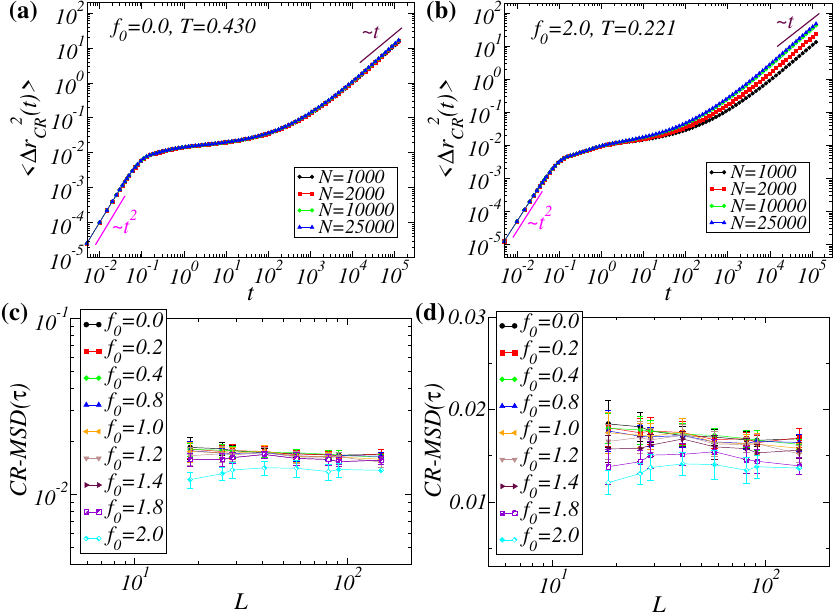}
	\caption{{\bf Cage-relative Correlation Function.} Cage-relative MSD for (a) passive and (b) active system. CR-MSD plateau plotted in log-log (c) \&  lin-log (d) clearly show no changes with increasing system size. This can be considered as a proof that the increase in MSD plateau is due to long wavelength modes in the system, which is absent in the cage-relative measurements. Error bars in the figure panels are measured by computing the standard deviation (SD) of fluctuations in various statistically independent simulations.}
	\label{fig:msdvsL_CM_merge}
\end{figure}

We now briefly study whether the long wavelength fluctuations affect the dynamics of active liquids at high enough temperatures where the effect of active force is much weaker than the thermal fluctuations, we have plotted the diffusivity ($D$) as a function of characteristic relaxation time, $\tau_\alpha$ (see Methods Section). Indeed, $D$ vs $\tau_\alpha$ plot \SK{in Fig. \ref{fig:MerminWagner}(f)} shows a power-law relation with exponent ($\kappa$) larger than $1$ ($\kappa \simeq 1.35$) at high temperatures and then at supercooled temperature regime $\kappa \simeq 0.80$ showing the well-known fractional Stokes-Einstein relation. The violation of SE relation at high temperature with $\kappa > 1.0$ again corroborates with the previous observation of the effect of long wavelength phonon fluctuations in 2D passive liquids including colloidal glasses in experiments \cite{Li2019}. 

Thus, effect of long wave length modes persists even for active liquids. In the inset of the same figure panel, we plot $D$ vs $\tau_\alpha$ after removing the long wavelength fluctuations by computing cage-relative methods as discussed in the Methods section. One sees the validity of SE relation with $\kappa \simeq 1$ at high temperature, this reinforces the presence of phonon-like (long wavelength modes) excitations in active liquids, even at high-temperature. Note that in \cite{Li2019}, it was claimed that in the overdamped Brownian limit, the effect of MWH fluctuations are completely suppressed in the liquid dynamics and thus one does not see any effect of MWH fluctuations in high temperature liquid in that limit. \SK{However we show later that MWH fluctuations in solids persist} even in the overdamped regime and activity enhances these fluctuations irrespective of detailed nature of activity.

\begin{figure*}[!htb]
        \includegraphics[width=.97\textwidth]{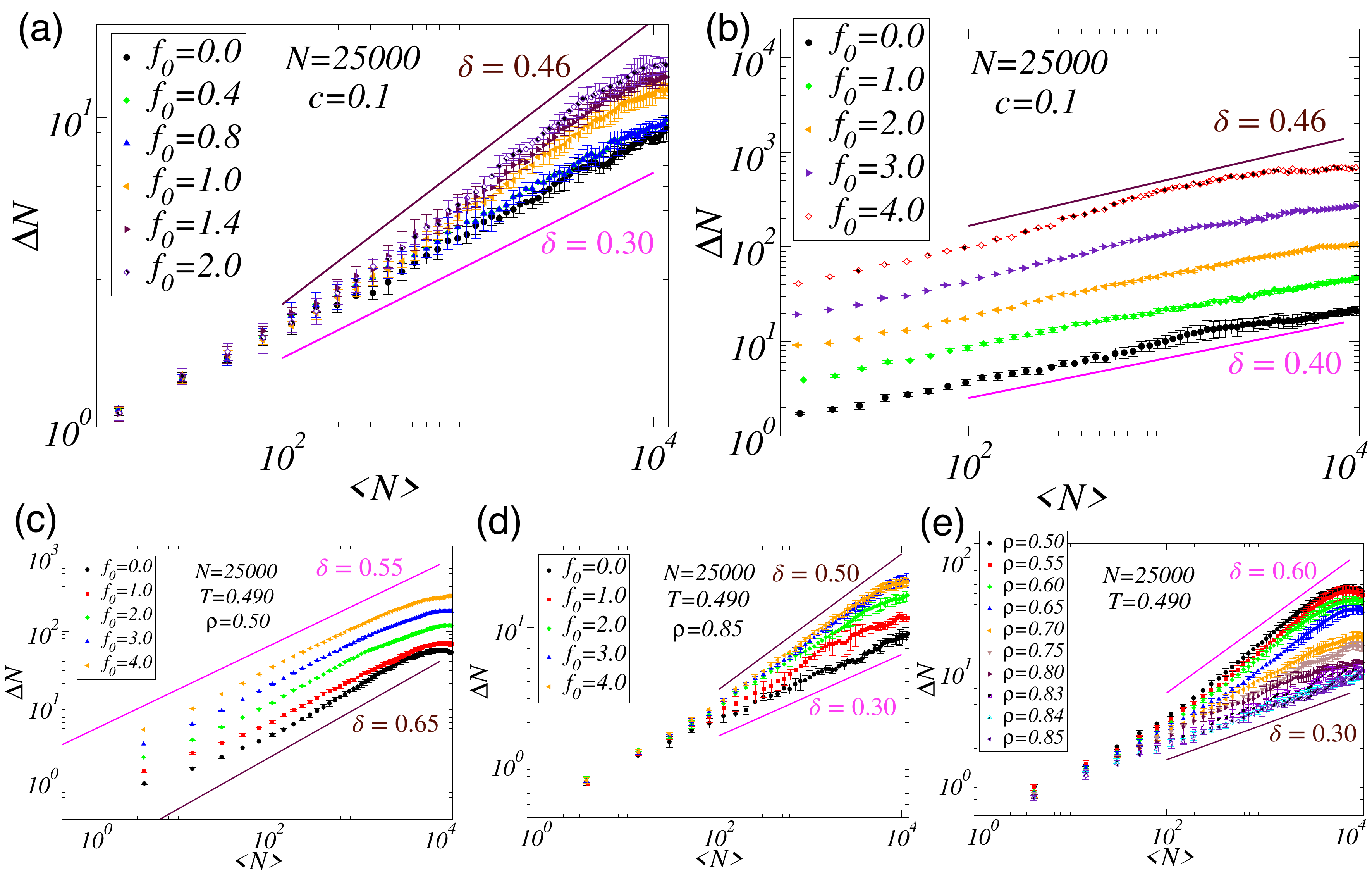}
        \caption{{\bf Giant Number fluctuation (GNF):} (a) In equilibrium, the fluctuations of the number of particles,  $\Delta N$ and average number of particles,  $\left<N\right>$ in a given sub-volume of linear size $l$ ($<L$) is related to each other as $\Delta N \propto \left<N\right>^{\delta}$ with $\delta=0.5$ in accordance with the central limit theorem.  Interestingly,  in disordered system this exponent $\delta$ tends to vary from $0.3$ to $0.46$ with increasing activity in 2dmKA model. (b) GNF for the 3DKA in the densed limit with changing activity, which shows the exponent is limiting towards the passive normal liquid case of $0.5$. (c) Shows the variation of $\delta$ from $0.65$ to $0.55$ with increasing activity in 2dR10 model with only repulsive inter-particle interactions at $\rho = 0.5$. The curves are shifted by a scale factor of ($1.5$) with respect to each other for better readability. (d) Shows that GNFs are absent in high density even with activity (see text for detailed discussion). (e) Shows for 2dR10 model, one sees the exponent $\delta$ to increase beyond $0.5$ with decreasing density indicating some presence of GNFs. But at very high density and low temperature the exponent ($\delta$) reaches $0.3$ for a disordered glassy system similar to 2dmKA model. Error bars in the figure panels are measured by computing the standard deviation (SD) of fluctuations in various statistically independent simulations.}
	\label{fig:GNF_plot}
\end{figure*}

To establish that the non-trivial effect of the increase in MSD plateau, is due to long-wavelength excitation modes on a firm ground, we have measured cage relative MSD (CR-MSD) to remove the effect of these modes from our measurements. In Fig.~\ref{fig:msdvsL_CM_merge}(a) and (b), we are showing the CR-MSD as a function of time given for passive and active glassy systems respectively. The CR-MSD plateau, defined as CR-MSD$(\tau)$, shows no divergence with system size for both the passive and active systems uniformly, as illustrated in Fig.~\ref{fig:msdvsL_CM_merge}(c) on a log-log scale and (d) on a linear-log scale. From this, one can be assured that the increase in MSD plateau is solely due to the long wavelength mode in the system, which is absent in the cage-relative measurements. Thus increasing MSD($\tau$) with the system size is a consequence of the MWH fluctuations even in non-equilibrium systems.

\vskip+0.1in
\noindent {\bf \large Possible effect of Giant Number Fluctuation (GNF): } GNF is a well-known characteristics of active systems in which particle density show disproportionately large fluctuations, deviating from typical equilibrium Gaussian statistics. In equilibrium, the standard deviation of the particle number ($\Delta N = \sqrt{\langle N^2 \rangle - \langle N\rangle^2}$) and its corresponding average particle number ($\left<N\right>$) in a sub-volume of  linear size $l$ ($<L$) is related by $\Delta N \propto \left<N\right>^{\delta}$, where $\delta = 1/2$ according to the central limit theorem. But the active systems are inherently in the out-of-equilibrium state and thus need not follow the same equilibrium relation; for systems with anisotropic particles, the exponent $\alpha$ is found to be greater than $0.5 $\cite{Narayan2007} in the polar order phase. Recently in  \cite{kuroda2023anomalous}, it was shown that GNF can arise in the fluid phase of active Brownian particles (ABP), where the polar order is absent at low number density of $\rho = N/V = 0.5$. In our system, we also observed the presence of GNFs in the low density limit but as we increase density GNF disappears. 

In Fig.~\ref{fig:GNF_plot} (a) we have plotted $\Delta N$ vs $\langle N\rangle$ for 2dmKA model system with $N = 25000$ particles.  The exponent $\alpha$ increases from $0.3$ to $0.46$ with increasing activity indicating the presence of hyper-uniform structures in these density regime. Thus for 2dmKA model at the studied density, although MSD plateau diverges with system size, it does not show any GNFs in the same activity limit. The results in 3D are very similar with exponent varying from $0.40$ to $0.46$ with increasing activity in 3dKA model (see Fig.\ref{fig:GNF_plot}(b)). These results are in complete agreement with the results reported in \cite{mitra2021hyperuniformity} for the same model. It was also pointed out that these class of systems show an effective hyper-uniform behaviour over a certain length scale as evidenced from the structure factor ($S(q)$ data), which shows a power-law like behaviour with wave vector ($q$) as $S(q) \sim q^{\beta}$ over a range of small $q$ but eventually saturates to a constant at much smaller $q$ \cite{ikeda2017large, mitra2021hyperuniformity}. The exponent $\beta$ is known to be connected to number fluctuations as $\Delta N \propto \left<N\right>^{(d-\beta)/2d}$, where $d$ is the spatial dimensions. Thus, $\delta = (d-\beta)/2d$. Thus, the hyperuniform behaviour found in our systems in both 2D and 3D are probably of this effective hyperuniform universality class showing a moderate suppressed density fluctuations.

\begin{figure*}[!htb] 
\centering
	\includegraphics[width=1.0\linewidth]{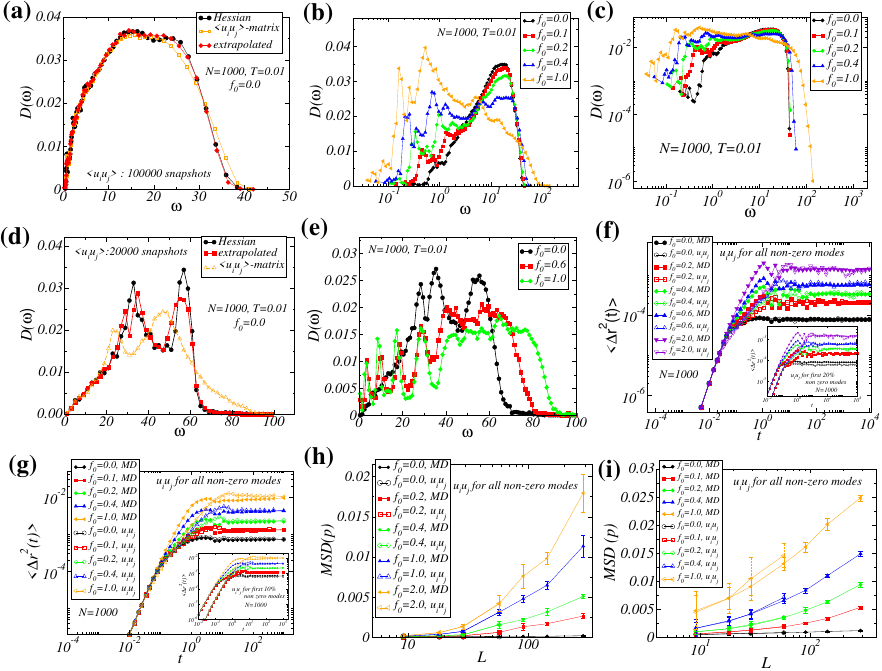}

	\caption{{\bf Effective Vibrational Density of States (vDoS) and Phonons:} (a) vDoS computed for passive systems for $N = 1000$ particles using exact diagonalization of the Hessian matrix (black circle), averaged displacement-displacement correlation matrix (orange square) and corrected via random matrix procedure (red diamond) (see text for details). The close agreement between these measurements suggests that the displacement-displacement correlation matrix method with random matrix correction is a robust method for the computation of vDoS. (b \& c) vDoS is computed using the displacement correlation matrix method for active systems with increasing activity. The clear appearance of small frequency ($\omega$) peaks in the vDoS with increasing activity signals the increasing dominance of phonon-like modes. The increasing weight of vDoS at small $\omega$ also indicates a jamming to unjamming scenario. (d) \& (e) vDoS computed for passive and active polycrystalline samples, respectively. (f) The measured MSD was compared with computed MSD from the correlation matrix. The excellent agreement suggests the validity of the effective dynamical matrix description of these active systems even at a significant degree of activity. (g) shows the same comparison for amorphous solids. Inset (f) and (g) highlight the importance of small $\omega$ modes in determining the plateau value of the MSD for all activities. (h) and (i) Comparison of plateau values vs system size, $L$, as obtained from MD simulations and from effective dynamical matrix description. This proves that dramatic Mermin-Wagner-Hohenberg (MWH) fluctuations in the active matter are due to the phonons, and thus, deviation of MWH theorem has to come from the details of the phonon dispersion relation. (f) \& (h) is for polycrystalline system, and (g) \& (i) is for amorphous solid. Error bars in the figure panels are measured by computing the standard deviation (SD) of fluctuations in various statistically independent simulations.}
	\label{fig:DoS}
\end{figure*}

To understand it further, we have taken another model of glass-forming liquids with purely repulsive interaction between particles (referred here as 2dR10 model; see Methods section for details) and varied the number density from low to high. The reason for choosing another model is that 2dmKA model at low density is known to show phase separation dynamics and we wanted to study a systems which is uniform at all studied density. In Fig.~\ref{fig:GNF_plot} (c), we show results for 2dR10 model at $\rho = 0.50$ with increasing activity, the exponent $\delta$ is nearly $0.55$, which shows mild GNFs. Fig.~\ref{fig:GNF_plot} (d) at high density, $\rho=0.85$ shows the exponent $\alpha$ to increase from $0.3$ to $0.5$ with increasing activity, which again does not show any sign of GNFs. While changing density we obtained exponent $\delta \simeq 0.3$ and $0.6$ in the high and low-density limit, respectively, as shown in Fig.~\ref{fig:GNF_plot} (e).  

By choosing $\delta = 0.3$ for the passive 2dmKA system as the baseline value, one might argue that an  increase in $\delta$ with activity will hint at GNFs, but our interpretation is that activity systematically reduces the degree of hyperuniformity within the studied activity limit as $\delta > 0.5$ is referred to as GNFs by definition. Also faster than logarithmic divergence in 2D and logarithmic divergence in 3D of Debye-Waller factor clearly highlight that  activity of the system enhances the MWH rather than other possible fluctuations.  We do not know any theoretical argument that suggests that GNFs can lead to different type of divergence in Debye-Waller factor in 2D and 3D system in the studied parameters regime.  Whereas our results can be very well rationalised using MWH fluctuations. Thus, we propose that GNFs do not play major role on the observed increase in MSD plateau with increasing system size, rather it is the MWH fluctuations that are dominant.

\vskip +0.1in
\noindent{\bf \large Effective Dynamical Matrix \& Phonons: }To investigate the long wavelength (phonon-like) behavior of the active crystals, and glasses, we have simulated both the systems at very low temperatures and employed a method to compute the effective dynamical matrix of the systems. This involved, calculating the displacement-displacement covariance matrix, as explained in Refs.~\cite{Chen2013, Still2014}. The covariance matrix ($\mathcal{C}$) is defined as $C_{ij}^{\mu\nu} = \langle u_i^{\mu}u_j^{\nu}\rangle$, where $u_i$ represents the displacement of the $i^{th}$ particle from its average position obtained by averaging over configurations from low-temperature MD data starting from a single minimum energy configuration, this configuration is found to remain close to the initial minimum configuration (minimum position). and $\mu$ or $\nu$ represent the spatial dimensions (such as $x$ or $y$ in 2D). The symbol $\langle\cdots\rangle$ denotes both the ensemble and time averaging. While exploring different approaches to computing the dynamical matrix, we found that the force-force correlation matrix method did not work well for active systems, as active force can not be derived from a potential as an active system lacks a Hamiltonian structure. However, this method proved to be very useful for passive systems (see detailed discussion in the \SK{Supplementary Note VIII,IX}). Therefore, we focused on the results obtained using the displacement-displacement covariance matrix, which, although computationally expensive, provided a good description of the system at a coarse-grained timescale. To obtain better convergence, we computed the correlation matrix at a timescale as small as a few molecular dynamics (MD) steps up to the largest timescale accessible. This is important to get good convergence at all frequencies. We performed computations on energy-minimized structures for the disordered systems, as well as polycrystalline solids, which we studied. This approach ensured that the matrix's eigenvalues were all positive, allowing us to validate the method's accuracy. Further details on this method can be found in the Method section and the \SK{Supplementary Note VIII,IX} . Note that the energy-minimized structures or the Inherent Structures (IS) are found to be the good minimum structures even in the presence of active forces. We have discussed these findings in the \SK{Supplementary Note XIII}.

In Fig.\ref{fig:DoS}(a), we show the vibrational Density of State (vDoS) obtained from the dynamical matrix $\mathcal{C}$ averaged over $64$ independent ensembles. The DoS obtained from the Hessian matrix $\mathcal{H}$ (see Methods for the definition) is labeled as "Hessian", and the one obtained using displacement-displacement correlation is labelled as "$\langle u_i u_j\rangle$ matrix".  Within Harmonic approximation, one can show $\mathcal{C} = (1/T)\mathcal{H}^{-1}$ (details in the \SK{Supplementary Note IX}), where $T$ is the temperature at which the displacement-displacement correlation matrix is obtained. The dynamical matrix does not depend on the particular choice of temperature as long as the temperature is low enough that the Harmonic approximation is a faithful description of the system, and during the simulation timescale, it does not escape out of the minimum.

\begin{figure*}[!htb] 
	\includegraphics[width=1.0\linewidth]{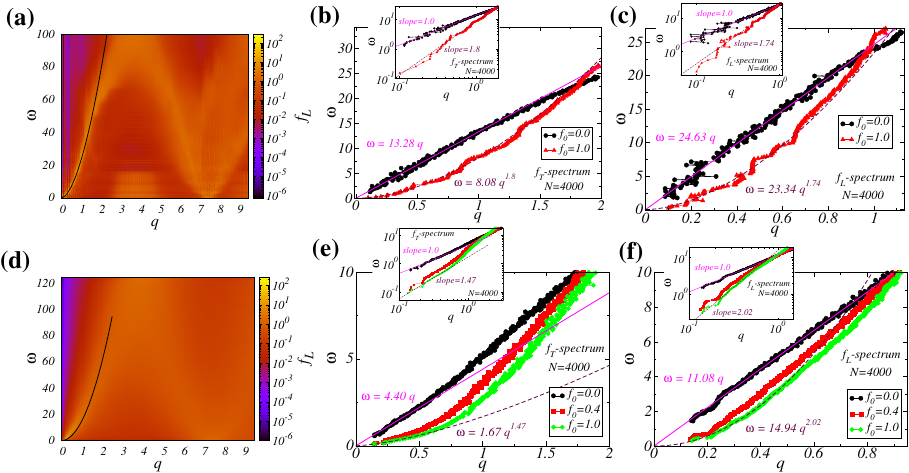}
	\caption{{\bf Effective phonon dispersion:} (a) Heat map of the $\omega$ vs $q$ for longitudinal spectrum of the polycrystalline samples with activity $f_0=1.0$. (b) shows the peak of heat map giving us the phonon dispersion relation of $\omega(q)$. The linear phonon dispersion relation for passive system is very clear and increasing non-linearity in active systems is also very evident, where for activity $f_0=1.0$ exponent is $\alpha \simeq 1.8$. Inset: log-log plot of the same. (c) the transverse spectrum of the polycrystalline samples for activity $f_0=1.0$ shows the dispersion exponent $\alpha \simeq 1.74$. (d) we show the heat map of the $\omega$ vs $q$ for longitudinal spectrum of the amorphous solid samples with activity f=1.0. (e) for amorphous solids samples with varying degree of activities. The spectrum is obtained for the longitudinal phonons and the inset shows the results in log-log plot. The exponent of the power law relation for activity $f_0=1.0$, $\alpha \simeq 1.47$. (f) shows the similar results but for transverse phonons with activity $f_0=1.0$ the exponent value is $\alpha \simeq 2.0$. Error bars in the figure panels are measured by computing the standard deviation (SD) of fluctuations in various statistically independent simulations.}
	\label{fig:Phonon_Dispersion}
\end{figure*}

In Fig.\ref{fig:DoS}(a), we present a corrected vDoS referred to as "extrapolated". We corrected the vDoS using a random matrix protocol, as described in detail in \cite{Chen2013, Still2014} (see \SK{Supplementary Note VIII}). In brief, we obtained the $\mathcal{C}$ matrix and calculated the eigenvalues ($\lambda = \omega^2$) of the matrix via numerical diagonalization procedure at various degrees of increasing ensemble averaging. Then, we re-estimated these eigenvalues via extrapolation to the infinite ensemble averaging limit based on results on random matrix theory, which suggest that the eigenvalues reach their true limiting values linearly with increasing averaging. The corrected vDoS matches well with the vDoS computed from the Hessian matrix as shown in panel A. We then used this procedure to obtain the dynamical matrix for active systems with increasing activity, as shown in Fig.\ref{fig:DoS}(b). We observed that the vDoS develops significant weight at smaller $\omega$, along with sharp peaks resembling the prominence of phonon-like modes at lower frequencies. The peaks become sharper and stronger with increasing activity. We represented the vDoS in a double logarithm plot in Fig.\ref{fig:DoS}(c) to highlight the similarity with the vDoS for jammed states approaching the unjamming transition. While it is true that with increasing activity, one approaches fluidization, the unjamming transition does not promote phonon-like excitations in the system. A detailed analysis is required to understand this similarity. In Fig.\ref{fig:DoS}(d), we show our results for polycrystalline samples without activity, and panel (E) shows the same with increasing activity, echoing the similar observation of enhanced phonon excitations with increasing activity.

\begin{figure*}[!htb] 
	\includegraphics[width=1.0\linewidth]{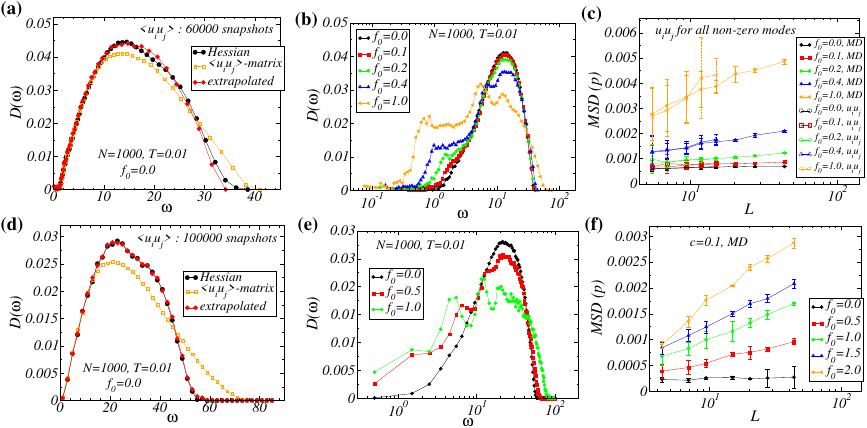}
	\caption{{\bf Effective vDoS and MSD plateau with activity in 3D:} (a) vDoS calculated using exact Hessian (black circle) and the effective dynamical matrix (orange square) for passive glassy system of system size $N=1000$ in 3D. Increasing the frame numbers leads to better convergence to the result from exact Hessian, which has been shown using the extrapolated vDoS (red diamond). Similarly results are shown in (d) for polycrystalline solid. (b) Using the effective dynamical matrix analysis for active glassy system, we observe that for higher activity the low energy frequency ($\omega$) modes show higher spectral weights. This proves that the activity has strong effect on the low frequency long-wavelength modes. Similar results are shown in (e) for polycrystalline solid. (c) MSD plateau shows the diverging behaviour with system size from MD simulation (bold), and MSD computed analytically using all modes of effective dynamical matrix also follows the same behaviour (empty). Strength of the logarithmic divergence increases with activity with no divergence in passive system. This is probably the first observation of logarithmic divergence of MSD plateau in 3D disordered system indicating a possible shift in lower critical dimensions ($d_l$). Similar results are shown in \SK{(f)} for polycrystalline system (see text for details). Error bars in the figure panels are measured by computing the standard deviation (SD) of fluctuations in various statistically independent simulations.}
	\label{fig:DoS_3D}
\end{figure*}

After reliable numerical computation of the vDoS, we wanted to verify the validity of this effective dynamical matrix in describing the dynamics by computing the MSD from the obtained eigenvalues and eigenvectors of the dynamical matrix, $\mathcal{C}$ as (see \SK{Supplementary Note VII} for the derivation)
	\begin{equation}
	\langle\Delta r(t)^2\rangle = \frac{K_B T}{N} \left[ \sideset{}{'}\sum_{a,i} (\mathbf{P}^a_i)^2 \frac{sin^2(\omega_a t/2)}{(\omega_a /2)^2}\right],
	\label{MSD_DoS}
	\end{equation}
where $\mathbf{P}^a_i$ is the $i^{th}$ component of the eigenvector $a$ and $\omega_a$ is the corresponding eigenvalue. In Figs.~\ref{fig:DoS}(f) and \ref{fig:DoS}(g), we compare the computed MSD from Eqn.~\ref{MSD_DoS} with the one obtained from the molecular dynamics (MD) simulation trajectories for polycrystal and amorphous solids, respectively. We observe a perfect match for the passive case, and a very good match for the active system. However, increasing activity strength shows a mismatch at intermediate timescales when the system transitions from the ballistic to the plateau regime. Despite this, both the short-time ballistic regime and the plateau regime are well-captured by the effective dynamical matrix. Although it is not immediately clear why there is a discrepancy at the intermediate timescale when the system transitions from the ballistic to the plateau regime, we believe that with increasing strength of the active forcing, the system takes a longer time to lose its memory of active driving in the particle displacement. Note that in dense disordered limit, both $f_0$ and $\tau_p$ play important role in determining the cage size which is shown to controls the dynamical behaviour of the active glasses \cite{debets2021cage}. 
At a longer timescale, the system will start to move towards the diffusive regime, and we believe that an effective dynamical matrix description will be a poor description. Thus, we expect that if we take only a few of the low-frequency modes of $\mathcal{C}$ and compute the MSD, we can correctly capture the plateau values for all activities. In the inset of Figs. \ref{fig:DoS}(f) and \ref{fig:DoS}(g), we show that the MSD computed using Eqn.~\ref{MSD_DoS} with only the first $10\%$ of the low-frequency modes indeed correctly captures the plateau value for all activities. These results suggest that the long-time behavior of the active system confined in a potential minimum can be accurately described by an effective dynamical matrix.

\vskip +0.1in
\noindent{\bf \large Effective Phonon Dispersion: }After establishing the effective phonon-like description of the active systems at low temperatures and thereby providing a strong evidence of phonon being the main reason behind the observed enhancement of MWH fluctuations, we want to understand the primary cause for the breakdown of the MWH theorem. Suppose we assume the basic mechanisms of MWH theorem arguments hold through. In that case, one expects that the phonon dispersion relation, which gives the dependence of $\omega$ on the wave vector $q$ must get modified due to the presence of active forces as follows:
	\begin{equation}
	\langle \Delta x^2\rangle \sim \int_{\frac{2\pi}{L}}^{\frac{2\pi}{\sigma}} \frac{qdq}{|\omega(q)|^2} \quad \sim L^{2\alpha -d},
	\end{equation}
	where we have assumed $\omega(q) \sim q^\alpha$. In the limit $\alpha \to 1$, we get back the usual logarithmic dependence, but for $2\alpha > d$, we will have a power-law divergence of the MSD with increasing system size. Thus, a non-linear phonon-dispersion relation in active systems can rationalize the observation. Microscopic understanding of why one expects a non-linear phonon dispersion relation is not clear immediately \SK{(see discussion on non-linear phonon dispersion section)}. 

\begin{figure*}[!htpb] 
	\includegraphics[width=1.0\linewidth]{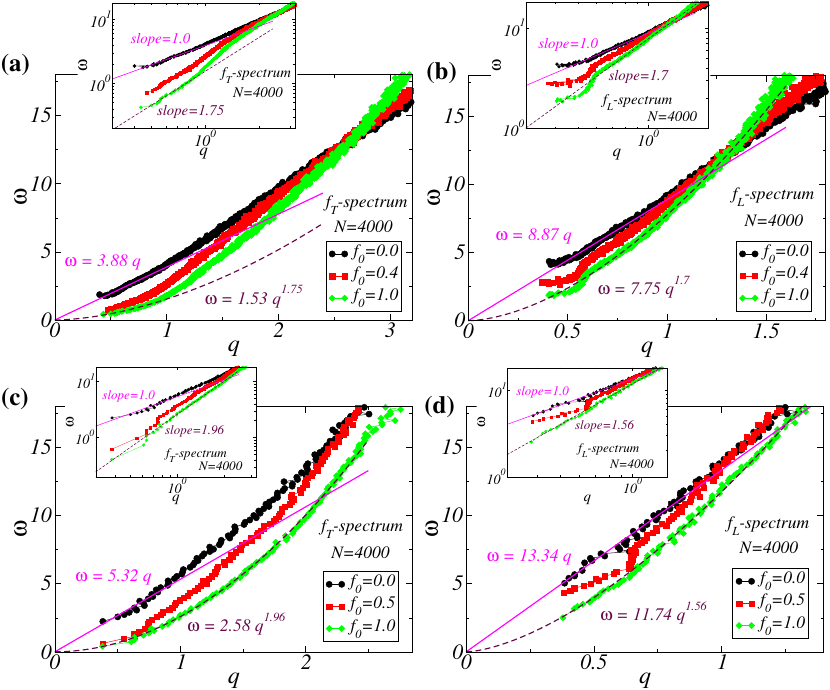}
	\caption{{\bf Effective phonon dispersion in 3D:} From the effective dynamical matrix we constructed the effective dispersion relation of the glassy and polycrystalline solid. For passive system dispersion relation in the $q \rightarrow 0$ shows the linear ($\omega \propto q$) dispersion relation for both the system for transverse as well as longitudinal spectrum. And for active glassy and polycrystalline solid it follows ($\omega \propto q^{\alpha}$) for active glassy and polycrystalline system. The exponent $\alpha$ for glassy system (a) transverse spectrum shows 1.75, (b) longitudinal spectrum 1.7. Similarly, for polycrystalline system (c) transverse spectrum shows 1.96 and (d) longitudinal spectrum 1.56. And all inset-plots is the log-log scale of its corresponding plots to the increasing nature of the exponent with activity. }
	\label{fig:Phonon_Dispersion_3D}
\end{figure*}

In Fig.\ref{fig:Phonon_Dispersion}(a), we show the heat map of the $\omega$ vs $q$ for longitudinal spectrum of the polycrystalline samples with activity $f_0=1.0$. In panel (b) shows the peak of the heat map, giving us the phonon dispersion relation of $\omega(q)$. The linear phonon dispersion relation for passive systems is very clear and increasing non-linearity in active systems is also very evident, Inset: log-log plot of the same plot shows the power-law exponent of the dispersion relation. For activity $f_0=1.0$, $\alpha \sim 1.8$. In panel (c), we show the dispersion relation for the transverse spectrum of the polycrystalline samples, the exponent $\alpha \sim 1.74$. 
Panel (d) shows the heat map of the $\omega$ vs $q$ for the longitudinal spectrum of the amorphous solid $f_0=1.0$ and panel (e) shows the dispersion with $\alpha \sim 1.47$ with inset showing the data in double logarithm. Panel (f) shows similar results but for transverse phonons with $\alpha \sim 2.0$. Results obtained for polycrystalline solids are in close agreement with the results obtained for amorphous solids and corroborate the robustness of an effective dynamical matrix description of the observation. Exponent $\alpha \sim 1.8$ for polycrystalline samples quantitatively describes the divergence of MSD plateau with $L$ as shown by the solid curve in Fig.\ref{fig:MerminWagner}(e). Similarly, if we choose $\alpha \sim 1.5$ for disordered solids, then we get $\delta \sim 1.0$, which is close to the exponent obtained from MD data. Further details are given in the \SK{Supplementary Note XI}.

\vskip +0.1in
\noindent{\bf \large Effective vDoS and MSD plateau in 3D:} To understand the effect of MWH fluctuations on dimension for both disorder glassy systems and for polycrystalline systems, we have done a similar study in 3D. In Fig.~\ref{fig:DoS_3D} (a), we have shown the vibrational density of states (vDoS) calculated using exact hessian (black circle) and the effective dynamical matrix (orange square) for a passive glassy system of system size $N=1000$ in 3D. Here, with increasing the frame numbers, the vDoS tends to converge to the exact Hessian curve, which has been shown using the extrapolated vDoS (red diamond). Again, in Fig.~\ref{fig:DoS_3D} (d), we can observe similar characteristics for polycrystalline solids. Now, taking a large enough number of frames, we can get the dynamical matrix for the active system as well. Fig.~\ref{fig:DoS_3D} (b) shows the vDoS computed from the effective dynamical matrix for an active glassy system; here, for higher activity, the weight of the low frequency ($\omega$) modes increases significantly with increasing activity as in 2D systems. Similar outcomes are observed in Fig.~\ref{fig:DoS_3D} (e) for polycrystalline solids with an increase in the spectral weight of the low-frequency modes. Thus in 3D also, activity plays a significant role in enhancing the low-frequency modes. Revisiting the Debye-Waller factor, which is proportional to the MSD plateau shown in Fig.~\ref{fig:DoS_3D}~(c) and (f) for glassy and polycrystalline systems, respectively,  we observe a diverging behavior with system size when activity is present and no divergence for passive case. The divergence of the MSD plateau increases with increasing strength of activity. Note that for all activity, one observes logarithmic divergence with system size (L). This is the first observation in the 3D system showing the logarithmic divergence of the plateau due to MWH fluctuations in active solids, which is induced solely by the presence of activity. This striking result is a novel observation, showing that MWH fluctuations can persist in higher dimensions $(d>2)$ for active glassy and polycrystalline systems.

\vskip +0.1in
\noindent{\bf \large Effective Phonon Dispersion in 3D:} Observed logarithmic divergence of MSD plateau with system size in 3D indicates \SK{the phonon dispersion relation to be independent of dimensionality}. To investigate the effect of the dimension on the effective dispersion, we explicitly computed the effective dispersion in 3D. This can be thought of as another validity test of the correctness of our proposed effective Hessian description. Using the eigenvectors of the dynamical matrix, we reconstructed the effective dispersion relation of the glassy and polycrystalline solid. We get back the linear dispersion relation ($\omega \propto q$) for a passive system in the $q \rightarrow 0$ limit for both the transverse as well as longitudinal spectrum of the system. For active glassy and polycrystalline solids, it follows $\omega \propto q^{\alpha}$ where, $\alpha>1$. The exponent $\alpha$ for the active glassy system for transverse and longitudinal modes are $1.75$ and $1.7$, respectively, shown in Fig~\ref{fig:Phonon_Dispersion_3D}(a) and (b). Similarly, for polycrystalline systems also, transverse and longitudinal modes are non-linear with $\alpha$ around $1.96$ and $1.56$, respectively shown in Fig.~\ref{fig:Phonon_Dispersion_3D}(c) and (d).

\vskip +0.1in
\noindent{\bf \large MWH Fluctuations with Active Brownian Particles (ABPs) :}
To establish the robustness of the observed results across a broad class of active matter systems both natural and synthetic active particles, we looked at the MWH fluctuations in active solids with active Brownian particles (ABPs). In a recent work in Ref.\cite{li2019long}, it was shown that passive colloidal particles (colloidal glasses) which perform Brownian motion, are not in overdamped limit as one finds Mermin-Wagner-Hohenberg (MWH) fluctuations in two dimensional (2D) colloidal assembly.  Thus, it is clear that active colloidal systems will be in the limit where inertia still plays important role and thus it is imperative to study active system with Brownian particles \SK{both in underdamped and overdamped limit} to see whether results obtained for active solids with RTPs are applicable for active solids with ABPs.  Also in \cite{wei2023reconfiguration},  it has been shown that RTPs embedded in a gel anneal the gel at an unprecedented rate which are very hard to achieve via usual thermal annealing, in close agreement with the results reported in \cite{Sharma2025}, where it was demonstrated that annealing by run-and-tumble active particles has a strong similarity with annealing via oscillatory shear.  These results suggests that active gels will also have enhanced MWH fluctuations in 2D and probably even in 3D systems. On the other hand, collection of cells in a monolayer will be in the overdamped regime. Thus MWH fluctuations in active solids with Brownian particles at various damping conditions will be very relevant for understanding experimental results in the field.

\begin{figure*}[!htpb]
	\includegraphics[width=1.0\textwidth]{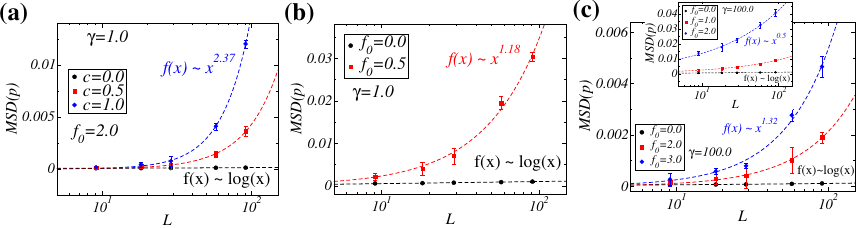}
	\caption{{\bf Mermin-Wagner-Hohenberg Fluctuations in Active Brownian Solids:} (a) Shows mean squared displacement (MSD) plateau for the polycrystaline system with $\gamma=1.0$ and $f_0=2.0$, exhibits a logarithmic divergence for passive Brownian particles and a faster-than-logarithmic divergence for active Brownian particles (ABP), with an exponent of $2.37$. We vary the active particle concentration $c$ from $0.0$ to $1.0$. (b) MSD plateau for 2dmKA model glass with $\gamma=1.0$ has an exponent of $1.18$ for active systems and usual logarithmic divergence for passive system. (c) In the extreme overdamped limit ($\gamma=100.0$), the divergence of the MSD for an active polycrystal has an exponent of $1.32$. In inset we show MSD plateau for passive and active Brownian particles in 2dmKA system in the extreme overdamped limit.}
	\label{fig:msd_plateau_gamma_100.0}
\end{figure*}

We performed extensive Brownian dynamics simulations and studied active Brownian particles (ABPs) in the dense glassy regime as well as in polycrystalline solid states in the damped and overdamped limit following protocol used in \cite{li2019long}. The equation of motion we use for simulating ABPs are 
	\begin{eqnarray}
	m \ddot{\mathbf{x}}_i &=& -\gamma \dot{\mathbf{x}}_i + \sum_{i \neq j = 1}^N \mathbf{f}_{ij} + f_0 \mathbf{n}_i + \zeta_i,\\
	\dot{\theta_i} &=& \xi_i
	\end{eqnarray}
	where each particle has mass $m$ interacting via Lennard-Jones potential as in 2dmKA model. The $i^{th}$ particle is subjected to a friction $\gamma$ and a thermal noise $\zeta_i$ with zero mean and variance $2k_B T \gamma \delta(t-t^\prime)$ that obeys fuctuation-dissipation relation. Each particle has a self-propulsion force $\mathbf{f}=f_0\mathbf{n}$ whose direction $\mathbf{n} \equiv \left( \text{cos } \theta, \text{sin } \theta \right)$ undergoes rotational diffusion with time evolution of $\theta_i$ is being described by an athermal noise $\xi_i$, with zero mean and correlation $\langle \xi_i(t) \xi_j(t^\prime)\rangle = 2 \tau_p^{-1} \delta_{ij} \delta(t-t^\prime)$. We fixed $\gamma = 1.0$ and $\tau_p = 20.0$ for the damped case. In Fig.\ref{fig:msd_plateau_gamma_100.0}(a) we show results for the same poly-crystalline model as before, with a number density of $\rho = 1.2$. For passive system (black circle) we see logarithmic divergence. Then we considered ABPs in two scenario: one where all particles were active (blue diamond) and another where only $50\%$ of the particles have active forcing (red square), for both the cases we fixed $f_0=2.0$. We see faster than logarithmic divergence in both cases and in the case where all the particles are active, the power-law exponent is $\sim 2.37$. In Fig.\ref{fig:msd_plateau_gamma_100.0}(b), we considered active Brownian particles in dense glassy regime. Again, we observed power-law divergence in the active case, where we set $f_0=0.5$ with all particles active. The exponent obtained is $1.18$. We could not set the active force to a larger value in the polycrystalline case as the particles were escaping from the sallow minimum.
	
\SK{Since MWH fluctuations are an equilibrium property, they do not depend on the microscopic dynamics of equilibrium solids. However, the manner in which the mean squared displacement approaches its asymptotic value does depend on the dynamics, but not the asymptotic value itself. Thus, it is much easier to observe signatures of these fluctuations in the dynamic properties with underdamped rather than overdamped dynamics. Hence, the question of possibility of observing these signatures with biologically relevant overdamped dynamics is an important question to address.} To study the system in the heavily overdamped limit, we fixed $\gamma=100.0$ and active particle concentration to be $c=1.0$ following \cite{li2019long} and studied poly-crsytal and glass with and without active forcing. We see power-law divergence for active systems (see Fig.~\ref{fig:msd_plateau_gamma_100.0} (c)), for poly-crystalline system we get exponent of $\sim 1.32$. For glassy systems we could not again increase the forces to larger values due to sallowness of the minima compared to the polycrystalline minima, but even there MSD plateau grows as $L^{1/2}$ in the large damping limit. \SK{In \cite{kuroda2023anomalous}, a fully damped ABP model was considered at zero temperature ($T = 0$) to show that MWH fluctuations becomes stronger than passive systems in the infinite persistence limit ($\tau_p \to \infty$).} These results clearly suggest that the stronger than logarithmic divergence of position fluctuations with systems size in 2D seem to be very generic result applicable to a broad class of active matter systems including many biological systems.

\begin{figure}[!htb] 
\centering
	\includegraphics[scale = 0.57]{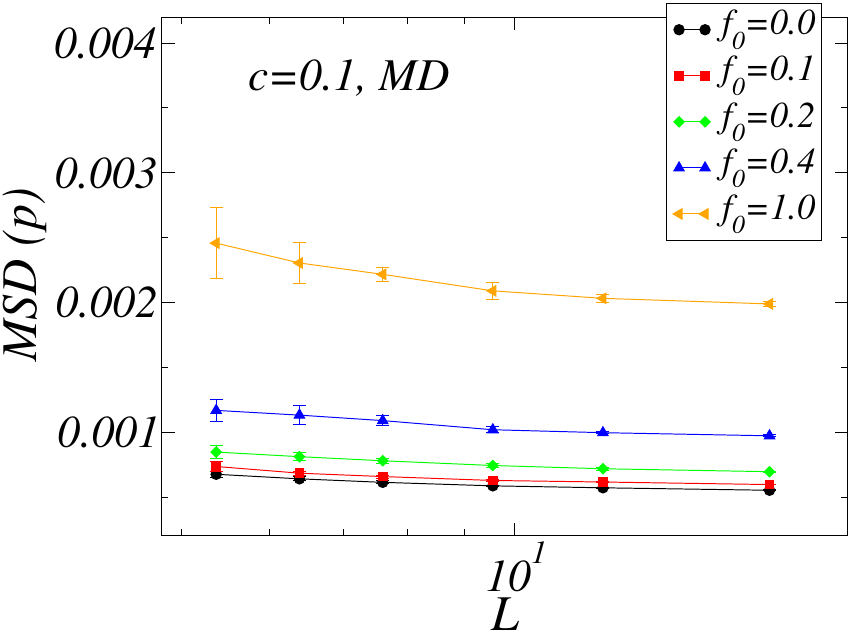}
	\caption{{\bf No MWH Fluctuations in 4D Active Solids:} Absence of MWH fluctuations in amorphous solids in four dimensions (4D). One can observe that the MSD plateau for the passive system decreases with increasing system size, reaching a constant independent of system size. For large activity, the MSD plateau also shows behavior similar to passive systems, albeit at a higher value, but no divergence with system size is observed. This shows that activity does not destabilize an active solid in 4D. Error bars in the figure panels are measured by computing the standard deviation (SD) of fluctuations in various statistically independent simulations.}
	\label{fig:4D_MSD}
\end{figure}

\vskip +0.1in
\noindent{\bf \large Shift in Lower Critical Dimension $d_l$:} 
Lastly, to understand whether the lower critical dimensions of system $d_l$ is shifted from $d_l =2$ in passive system to $d_l = 3$ in active solids or it is an intrinsic property of active systems at any dimensions to show anomalous dynamical properties, we also simulated the Kob-Andersen model in four spatial dimensions (4D) at low-temperature $T=0.01$ at density $\rho = 1.2$. In Fig.~\ref{fig:4D_MSD}, one can observe that the MSD(p) for the passive system decreases with increasing system size, reaching to an asymptotic value that is independent of system size. For large activity, the MSD(p) also decreases with system size from its smaller system size value before saturating to a constant in the large system size limit. Thus, in 4D system, the fluctuations seem to be normal, proving beyond doubt that the lower critical dimensions of the system increased from $2$ to $3$ in active solids with RTP-type active forcing.  This is the first time we can show that the activity can shift the lower critical dimension ($d_l$) of the system to $3$ in the studied activity limit. To gain deeper insight, we followed the approach by \cite{Henkes2020} and conducted a similar analytical calculation to compute the mean-squared displacement for an active system within linear response formalism. In their work, Henkes et al.  \cite{Henkes2020} used continuum active linear elasticity and a normal modes formalism to understand the results obtained in epithelial cell monolayers experiments. This allowed them to calculate velocity correlations over distance analytically. Building on this approach, we also derived an analytical expression for the mean squared displacement (MSD) in the following section.

\vskip+0.1in
\noindent{\bf \large Discussion on non-Linear Phonon Dispersion:} Although we are not able to offer a microscopic theory to explain the non-linear phonon dispersion relation, an interesting connection to vDoS in fractals is worth considering. If we assume that the underlying structures that support the phononic excitations in these non-equilibrium solids are fractals with fractal dimensions less in $2$ in 2D and then it can be argued that vDoS on these fractals $\sim \omega^{1/3}$ \cite{Orbach1982} (also see \cite{YakuboNakayama1987a, YakuboNakayama1989, RMP_Orbach1994} on percolating network in $d=2$). Assuming such a vDoS, phonon dispersion relation for a finite size system with linear size $L$ will read as $\omega \sim q^{3/2}$ with $q$ being the wave vector. For a general non-linear phonon dispersion of the form $\omega(q) \sim q^{\alpha}$, vDoS will have the following form
\begin{equation}
\mathcal{D}(\omega) \sim \frac{L^2}{\pi} \left ( \frac{\omega}{C_s}\right)^{\frac{2-\alpha}{\alpha}}.
\end{equation} 
See \SK{Supplementary Note XII} for detailed derivation. If we now assume that $\alpha \simeq 1.5$, then we get $\mathcal{D}(\omega) \sim \omega^{1/3}$ and cumulative vDoS going as $\omega^{4/3}$. In \SK{Supplementary Note XII}, we showed that the cumulative vDoS  $P_c(\omega)\sim \omega^2$ for the passive system, whereas the active system shows much weaker dependence on $\omega$. Thus, an enhanced vDoS at a small frequency could be a possible explanation for the non-linear phonon dispersion relation. Our effective Hessian calculation suggests that the phonon modes remain very similar in the presence of active forces but a lot many modes became de-localized as well as their frequency became smaller. Although the question of underlying fractal networks controlling the vibrational properties of these active solids remains to be validated but enhanced force-chain networks in the presence of active forces and their role in determining the mechanical properties of these solids will be interesting to investigate.

\vskip +0.1in
\noindent{\bf \large Breakdown of existing Linear response formalisms:} In \cite{Henkes2020}, the long wavelength modes for monolayer of soft, self-propelled agents packed densely using a dynamical matrix approach were calculated. Using the same framework, we computed the mean-squared displacement (MSD) for the same damped active Brownian system. Below, we briefly discuss the salient parts of the calculation for \SK{the completeness of the discussion in the subsequent sections}. The equation of motion for particles can be written as 
\begin{equation}
	\zeta \dot{\mathbf{r}}_i = \mathbf{F}^{\text{act}}_i +  \mathbf{F}^{\text{int}}_i
	\label{eom}
\end{equation}	
where, $\zeta$ is the friction coefficient, $\mathbf{F}^{\text{act}}_i$ is the net active force $\mathbf{F}^{\text{int}}_i$ is the force on $i$-th particle exerted by its neighbours.

Next, Linearizing the interaction forces in the vicinity of an energy minimum, $\{ \mathbf{r}_i^0\}$ by introducing $\delta \mathbf{r}_i = \mathbf{r}_i -\mathbf{r}_i^0$. Eq.~(\ref{eom}) becomes
\begin{equation}
\zeta \delta {\dot{\mathbf{r}}_i} = \zeta v_o \hat{n}_i - \sum_j \mathbf{K}_{ij} \cdot \delta \mathbf{r}_j
\label{eom_proj}
\end{equation}
where, $v_0$ is the active velocity and $\mathbf{K}_{ij} = \frac{\partial^2 V(\{\mathbf{r}_i\})}{\partial \mathbf{r}_i \partial \mathbf{r}_j}$ is the dynamical matrix. Now $\mathbf{K}_{ij}$ has $2N$ number of normal modes $\mathbf{\xi}^\nu$, with positive eigenvalues $\lambda_\nu$ and we can project Eq.~(\ref{eom_proj}) onto it, and obtain
\begin{equation}
	    \zeta \dot{a_\nu} = -\lambda_\nu a_\nu + \eta_\nu
	    \label{normal_mode}
\end{equation}
where, $a_\nu = \sum_i \delta \mathbf{r}_i \cdot \mathbf{\xi}_i^\nu$ and active force are projected onto the modes, $\eta_\nu = \zeta v_0 \sum_i \hat{n}_i \cdot \mathbf{\xi}_i^\nu$ is the active persistence noise with $\langle \eta_\nu(t) \eta_\nu^\prime(t^\prime)\rangle = C(t-t^\prime)\delta_{\nu\nu^\prime}$. The correlation function $C(t) =\frac{\zeta^2v_0^2}{2}e^{-t/\tau_p}$ with $\tau_p$ being the persistence time. Now, we can integrate Eq.~(\ref{normal_mode}) and obtain the moments of $a_\nu$ and from there we can compute the MSD as 
\begin{equation}
	MSD = \frac{L^2}{2\pi^2} \int d^2 \mathbf{q} \sum_\nu \frac{\zeta v_0^2 \tau_p}{2 \lambda_\nu \left( 1 + \frac{\lambda_\nu}{\zeta} \tau_p \right)} || \mathbf{\xi}_\nu (\mathbf{q}) ||^2
	\label{msd_fourier}
\end{equation}
where, $\mathbf{\xi}_\nu(\mathbf{q})$ is Fourier transform of eigenvector $\xi_i^\nu$.

Next we turn to continuum limit to write Eq.~(\ref{eom_proj}) as
	\begin{equation}
		\zeta \dot{\mathbf{u}}(\mathbf{R}) = \zeta v_0 \hat{n}(\mathbf{R}) - \sum_{R^\prime} \mathbf{D}(\mathbf{R} - \mathbf{R}^\prime)\mathbf{u}(\mathbf{R}^\prime)
		\label{eom_continuum}
	\end{equation}
where, $\mathbf{u}(\mathbf{R})$ denotes the elastic deformations from the equilibrium position $\mathbf{R}$, and $\mathbf{D}(\mathbf{R} - \mathbf{R}^\prime)$ is the continuum dynamical matrix.  Fourier transforming Eq.~(\ref{eom_continuum}) 
and writing $\mathbf{\tilde{F}}^{act} (\mathbf{q}, \omega) = \zeta v_0 \int d^2 \mathbf{r} \int_{-\infty}^{\infty} dt \hat{n}(\mathbf{r}, t) e^{i \mathbf{q}\cdot \mathbf{r} + i \omega t}$, one can calculate the correlation of active noise, $\left< \mathbf{\tilde{F}}^{act} (\mathbf{q}, \omega) \cdot \mathbf{\tilde{F}}^{act} (\mathbf{q^\prime}, \omega^\prime) \right>$. If we then write transverse and longitudinal modes of $\left<\mathbf{\tilde{u}}(\mathbf{q}, \omega) \cdot \mathbf{\tilde{u}}(\mathbf{q}^\prime, \omega^\prime)\right>$ in terms of active noise, we will be able to obtain the final expression of MSD for large $L$ in real space and time as (see \SK{Supplementary Note XIV} for the detailed derivation)
\SK{
The generalised mean squared displacement in d-dimension is given by,
\begin{widetext}
\begin{equation}
	\left<|\mathbf{u}(\mathbf{r},t)|^2\right> = \left(\frac{1}{2\pi}\right)^{d/2} a^{d} \ \frac{(v_0 \tau)^2}{2} \frac{1}{\Gamma(d/2)}
	\ \int dq \ \frac{q^{d-1}}{q^2} \left[\frac{1}{\xi_L^2(1+\xi_L^2 q^2)}
	+ \frac{1}{\xi_T^2(1+\xi_T^2 q^2)}
	\right]
\end{equation}
\end{widetext}
where, longitudinal and transverse length scale is given by, $\xi_L = \left[(B+\mu) \tau/\zeta\right]^{1/2}$ and $\xi_T = \left[\mu \tau/\zeta\right]^{1/2}$. For $d = 2$, we get the analytical expression for the MSD as 
\begin{widetext}
\begin{align}
	\left<|\mathbf{u}(\mathbf{r},t)|^2\right> &= 
	\frac{1}{4\pi} a^2 v_0^2 \zeta \tau \cdot \frac{1}{2(B+\mu)} \left[ 2 \ln(L/a) - \ln \left(1+\left(\frac{2\pi}{L} \xi_L\right)^2 \right) + \ln \left(1+\left(\frac{2\pi}{a} \xi_L\right)^2 \right) \right] \notag \\
	& + \frac{1}{4\pi} a^2 v_0^2 \zeta \tau \cdot \frac{1}{2\mu} \left[ 2 \ln(L/a) - \ln \left(1+\left(\frac{2\pi}{L} \xi_T\right)^2 \right) + \ln \left(1+\left(\frac{2\pi}{a} \xi_T\right)^2 \right) \right].
\end{align}
\end{widetext}
If we take $L \rightarrow \infty$ of the above equation, then we get 
\begin{align}
	\left<|\mathbf{u}(\mathbf{r},t)|^2\right> = 
	\frac{1}{4\pi} a^2 v_0^2 \zeta \tau \cdot \left[ \frac{1}{(B+\mu)} + \frac{1}{\mu}\right] \ln(L/a)
\end{align}
}
This result predicts a logarithmic divergence with system size $L$ for active systems as well (see \SK{Supplementary Note XIV}), which are in contrast with our results.  \SK{However, if one takes extreme activity limit by making persistence time $\tau_p \to \infty$, one finds the MSD plateau to diverge as $\left<|\mathbf{u}(\mathbf{r},t)|^2\right> \sim L^{4-d}$, suggesting the lower critical dimensions to shift to $4$ as also claimed in \cite{kuroda2023anomalous,HIkeda_LCD}. Note that in the limit $\tau_p\to\infty$, both lognitudinal and transverse correlation lengths of the active elasticity theory diverge as $\sqrt{\tau_p}$ \cite{kuroda2023anomalous}. At finite persistence time, these correlation lengths will be finite and it will predict a crossover behaviour beyond these length scales. Notably, it will predict in 3D, $\left<|\mathbf{u}(\mathbf{r},t)|^2\right> \sim L$ if $L<\xi_L$ or $\xi_T$ and become constant beyond this crossover length scale similar to passive systems. In 2D, it will predict $\left<|\mathbf{u}(\mathbf{r},t)|^2\right> \sim L^2$ for $L<\xi_L$ or $\xi_T$ and $\log{L}$ beyond these correlation lengths. In contrast, our results in 3D at finite persistence time suggest $\left<|\mathbf{u}(\mathbf{r},t)|^2\right> \sim \log{L}$ with the lower critical dimensions to be $3$ rather than $4$. Although, our initial results in 3D, suggests logarithmic divergence of MSD plateau with increasing system size over the studied $\tau_p$ range (see Supplementary Fig.20b for details), it will be very interesting to do a systematic study with varying persistence time over a much larger window including $\tau_p \to \infty$ to test some of these theoretical arguments.} Thus, we believe that the existing theories based on continuum active linear elasticity and the normal modes formalism is not sufficient to explain power-law divergence and logarithmic divergence of MSD plateau with system size in 2D and 3D respectively \SK{ in different model active solids over a wide range of activity parameters}. Theoretical formalism that can incorporate a detailed renormalization of the dynamical matrix along with the phonon dispersion relation is essential to understand the results presented in this manuscript and we hope that our work will inspire new theoretical approaches to develop a correct theory of active solids.

\vskip +0.1in
\noindent{\bf \large Discussion:}
We performed extensive computer simulations to study the effect of long wavelength Mermin-Wagner-Hohenberg like fluctuations in polycrystalline solid and glasses in liquid and solid regimes, and observe the dimensionality effect on the active systems. We found that active forces modelled as run-and-tumble particles (RTP) \SK{and active Brownian particles (ABPs)} significantly enhance the long wavelength density fluctuations in these systems. These long-wavelength density fluctuations also affect liquid dynamics at high-temperatures, similar to the results obtained in passive systems\cite{li2019long}. For an active glassy system in 2D, MWH fluctuations cause a power-law divergence of the Debye-Waller (DW) factor with system size, $L$, \SK{whereas in 3D active solids, one observes logarithmic divergence which is not reported in the literature before}. The exponent increases with increasing activity, universally in both crystalline and disordered solids. 

Once long wavelength fluctuations are removed from the measurement by calculating cage-relative MSD, one observes the system size divergence of the MSD plateau to vanish. This proves that the observed divergence with system size is solely due to the collective motion of the long wavelength mode, which gets enhanced in the presence of activity. Moreover, by performing finite size scaling analysis of the glassy system with changing interactions and density in the presence of activity, we are able to show that the non-trivial fluctuations in the 2D \SK{and 3D} active system do not stem from giant number fluctuations (GNFs) at the studied density range. Following \cite{Henkes2020}, we calculated the displacement fluctuation in active systems within linear response and active elasticity theory and found that it predicts usual logarithmic divergence even for active systems. This result suggests that existing theoretical formalisms of active solids are not adequate to explain our current observations. Our results indicates that assumption of dynamical matrix being the same as the passive system in the presence of active colour noise is probably at fault.

We proposed an effective dynamical matrix description of the observation by computing the effective Hessian matrix from displacement-displacement covariance matrix and performing exact diagonalization to compute the eigenvalues and eigenvectors. The obtained vibrational density of states (vDoS) matched very well for the passive systems after systematically correcting for convergence issues. vDoS results obtained following the same protocols led to the observation of enhanced phonon modes at lower frequencies with increasing activity, as evident from the appearance of sharp peaks in the distribution at small frequencies. The validity of this effective description was verified by computing the mean squared displacement (MSD) of particles in both crystals and amorphous solids using the detailed information of eigenvalues and eigenvectors of the effective Hessian matrix, showing close agreement with MD simulations results. Subsequently, to shed more light on the power-law and logarithmic divergence of DW factors 2D and 3D in crystalline and amorphous solids respectively, we have computed the effective phonon dispersion relation and showed that in passive systems, one gets back the usual linear dispersion as $\omega(q) \sim q$, whereas in active solids, this relation becomes non-linear as $\omega(q) \sim q^\alpha$, with exponent $\alpha \sim 1.8$ in polycrystalline solids and around $1.47-2.02$ in amorphous solids with exponent increasing systematically with increasing activity in both 2D and 3D. These results qualitatively rationalize the observed power-law and logarithmic divergence of DW in crystalline and amorphous solids in 2D and 3D respectively as well as no divergence in 4D solids.

For an out-of-equilibrium system, one does not expect the MWH theorem to hold like the equilibrium system, and there is plenty of work present in the literature, where for an out-of-equilibrium scenario the MWH theorem breaks down, i.e., the system shows a long-range crystalline ordered state in 2D. In a recent study \cite{Galliano2023}, it has been shown that due to active noise, the 2D system attains long-range crystalline order, which denotes that the MWH fluctuation in the system is suppressed. This result is in agreement with the out-of-equilibrium disorder-to-order transition observed in active systems \cite{Vicsek1995, Toner1998, Palacci2013}. Unlike these systems, our systems show MWH fluctuations to persist with a great strength due to the colored noise statistics of the self-propelled particles (SPPs) present in the system. Another recent studies \cite{Umang2023, Sharma2025} suggest that the RTPs get strongly coupled to the global and local shear modes of relaxation, which are collective in nature, leading to faster annealing similar to the oscillatory shear response of these solids. We believe that if active driving efficiently gets coupled to the local shear modes, then it will also lead to enhanced phonon-like fluctuations in the system. It will be interesting to investigate the effective phonon description in systems where activity suppresses the long wavelength density fluctuations as in Ref.\cite{Galliano2023}. Such studies may shed insight into whether the phonon dispersion changes consistently in those systems.

Our study reveals for the first time that run-and-tumble particles (RTPs) and active Brownian particles (ABPs) can disrupt the stability of solids in a fundamentally different way, causing destabilization even in three-dimensional solids, leading to an increase in the lower critical dimensions from $d_l = 2$ to $d_l = 3$. This finding echoes the well-known arguments by Imry and Ma \cite{Imry1975}, which discuss how quenched disorder can alter the lower critical dimension in magnetic systems. Establishing a potential connection between magnetic systems with quenched disorder and active solids with RTPs or ABPs is an exciting prospect. In \cite{HIkeda_LCD},  it was predicted that lower critical dimensions can increase when the driving noise is positively correlated in space and time and it can decrease when it is anticorrelated.  Lower and upper critical dimensions were calculated analytically for the spherical model $\mathcal{O}(n)$ in the limit $n\to \infty$, in presence of correlated noise.  It will be indeed very important to find out whether any such mechanism is at play in our system.  Naively,  both RTPs and ABPs have temporal correlation in the noise statistics but no spatial correlation,  although an emergent dynamical correlation can not be ruled out at this point. Detailed investigation along some of these directions will be very helpful to improve our understanding of the microscopic mechanisms behind the amplified fluctuations observed in these systems.  Finally,  the validity of our results across various active matter model systems and at different damping conditions suggests that these findings could be applicable to a wide range of active matter systems, both synthetic and biological matters.

\section*{Methods}
\noindent{\bf \large Models \& Simulation Details: }In this work, we have studied the dynamics of crystalline solid and model glass-forming liquids. The first glass-forming liquid model is the modified Kob-Anderson (2dmKA) Binary model in 2D with a composition ratio of $65:35$ with the other details of the potential same as that of the Kob-Anderson model \cite{Das2017} (see \SK{Supplementary Note I} for details). This particular composition ratio ensures that there is no tendency for the system to form local crystalline orders. For the polycrystalline system, we have taken a mono-atomic particle system with energy and the diameter of the particles to be $1.0$ in LJ unit, and the interaction potential is smoothed LJ potential, same as the other two glass-forming models. We have performed simulation with number of particles in the range $N \in [400, 10^5]$ in this work. We ran $32$ statistically independent ensembles for all the systems except the few large ones ($25000 - 10^5$); we have taken $8$ ensembles for these system sizes. We used three-chain No\'se-Hoover thermostat to perform NVT simulations \cite{Martyna1992}. Polycrystal model in 3D is the same as the 2d polycrystal model but simulated in three dimensions. 3dKA and 4dKA models for disorder systems have the same interaction parameters as the 2mdKA but simulated in three (3D) and four dimensions (4D) with a binary particle ratio of $80 : 20$ (large : small), respectively.

\vskip +0.1in
\noindent{\bf \large Implementation of Activity: }The activity in the system is introduced in the form of run and tumble particle (RTP) dynamics \cite{Mandal2016, Kallol2021, Dey2022}, where the dynamics of the active particles can be tuned using three parameters such as c, $f_0$, $\tau_p$. The equation of motion is given by
\begin{eqnarray}
\dot{\mathbf{r}}_i&=&\frac{\mathbf{p}_i}{m}\nonumber\\
\dot{\mathbf{p}}_i &=& -\frac{\partial \Phi}{\partial \mathbf{r}_i}+n_i\mathbf{F}^A_i,
\end{eqnarray}
with $\mathbf{r}_i$ and $\mathbf{p}_i$ being the position and momentum vector of $i^{th}$ particle, $n_i$ is the active-tag which take values $0$ or $1$ depending on whether the particles is passive or active, $\Phi$ is the inter-particle potential, and $\mathbf{F}_i^A$ is the active force. The active force on the $i^{th}$ particle in 2D can be written as,
\begin{equation}
\mathbf{F}^A_i = f_0 (k^i_x \hat{x}+k^i_y \hat{y}),
\end{equation}
where the active direction of the $i^{th}$ particle is $k^i_{\alpha}$ for $[\alpha \in x,y]$ is chosen from $\pm1$. The number of total active particles is taken to be an even number to maintain the total active momentum to be zero along all directions independently, i.e., $\sum_{\alpha,i} k^i_{\alpha}=0$. The basis set in 3d and 4d would be $[\alpha \in x,y,z]$ and $[\alpha \in x,y,z,w]$ respectively.
For setting up the reference temperature, we kept $\tau_p = 1$, active force magnitude $f_0$ is selected from $0.0 - 2.0$ for 2dmKA by fixing concentration $c=0.1$. 

\vskip +0.1in
\noindent{\bf \large Active P\'{e}clet number (Pe): }It is often useful to define the degree of activity in a non-dimensional form. For active system the P\'{e}clet number (Pe) is such a quantifier to get the amount of activity in the system. Pe is a non-dimensional number which is the ratio of the transport rate due to advection (activity) and diffusivity (D) of the system. Typically for any system Pe can be defined as $Lv/D$. But for active system it can be modified such that it can quantify the distance travelled by a free particle due to activity in terms of velocity, this P\'{e}clet number for active system is defined as, 
\begin{equation}
	Pe= \frac{\tilde{v}_A}{\tilde{v}}=\frac{f_0 \cdot \tau_p}{\gamma \cdot \sigma_{AA}},
	\label{PeEq}
\end{equation}
where $\tilde{v}_A$ is the persistent velocity of the active particles given by ($f_0 \cdot \tau_p$) and $\tilde{v}$ is the velocity of the particle due to inertial (velocity) relaxation of the system. The inertial relaxation time ($\tau_{\gamma}$) of the system is $\gamma ^{-1}$ and $\sigma_{AA}$ is the diameter of the large particle \cite{Fily2014,Mandal2020a,Kuroda2023}.

For the deterministic molecular dynamics simulation using (Nose-Hoover) thermostat, the damping term ($\gamma$) is not present. If we use Einstein relation of fluctuation-dissipation theorem $\gamma=K_B \cdot T/D$, then the P\'{e}clet number in our case will read as 
\begin{equation}
	Pe= \frac{f_0 \cdot \tau_p}{(K_B \cdot T/D) \cdot \sigma_{AA}}.
\end{equation}
where, Boltzmann constant $K_B$ and $\sigma_{AA}$ is 1 in reduced LJ unit. If we consider that the damping term is $\gamma \simeq 1$, then the persistent length ($l_p$) of the active system will be given by, $l_p\propto f_0 \cdot \tau_p^2$ and as we have taken $\tau_p = 1$ and varied $f_0$ from $0$ to $2.5$ in this study,  the persistence length will be roughly $1-2$ particle diameter for the largest activity at low densities but due to dense crowded environment,  the caging length remains $\sim 0.7\sigma_{AA}$. There are studies which suggests that in dense disordered systems,  caging length controls the dynamical behaviour of the active glasses \cite{debets2021cage}.  Note that the caging length of around $0.7\sigma_{AA}$ is not very different from the persistence length of a cell (around one cell diameter) for an assembly of cells.  For ABPs,  Pe is well defined via Eq.\ref{PeEq}. 

\vskip +0.05in
\noindent{\bf \large Correlation Functions: }To compute two-point density-density correlation, we have considered a simpler form of overlap correlation function Q(t), such as
	\begin{equation}
	Q(t) = \frac{1}{N} \sum^N_{i=1} w(|\mathbf{r}_i(t)-\mathbf{r}_i(0)|).
	\end{equation}
	where $w(x)$ is a window function, which is $1$ for $x<a$ and 0 otherwise. $\mathbf{r}_i(t)$ position of the $i^{th}$ at time $t$. Here parameter `a' is chosen to remove the vibration due to the caging effect. We chose $a = 0.3$. Relaxation time, $\tau_{\alpha}$, which is defined as $\langle Q(t=\tau_{\alpha})\rangle = 1/e$, where $\langle\cdots\rangle$ means ensemble average. The system is equilibrated for $150\tau_{\alpha}$, and then we run for $150\tau_{\alpha}$ in all our measurements. 

Four-point correlation function can be measured from the fluctuation of the two-point correlation function Q(t), which is defined as,
	\begin{equation}
	\chi_4(t) = N \left[\left<Q(t)^2\right> - \left<Q(t)\right>^2\right].
	\end{equation}

\vskip +0.1in
\noindent{\bf \large Mean Squared Displacement (MSD): }The mean squared displacement (MSD) is defined as
	\begin{equation}
	\langle \Delta r^2(t)\rangle = \frac{1}{N}\sum_{i=1}^{N} |\mathbf{r}_i(t) - \mathbf{r}_i(0)|^2, 
	\end{equation}

The diffusion constant $D$ is computed from the slope of MSD vs t at long timescale using the relation \SK{$\langle \Delta r^2(t)\rangle = 2dDt$ in d dimensional system}. 

\vskip +0.1in
\noindent{\bf \large Cage-relative Displacement: }
The cage-relative (CR) displacement \cite{Mazoyer2009, Vivek2017, Illing2017} of the individual particle $i$, is defined as  
	\begin{equation}
	\mathbf{r}_{i,CR}(t) = \left[ \mathbf{r}_i(t) - (\mathbf{r}_{i,nn}(t) - \mathbf{r}_{i,nn}(0) ) \right]
	\end{equation}
	where $\mathbf{r}_{i,nn}(t)$ is the position of the center of mass of $N_{nn}$ nearest neighbours of $i^{th}$ particles. Again it is defined as,
	\begin{equation}
	\mathbf{r}_{i,nn}(t) = \frac{1}{N_{nn}} \sum^{N_{nn}}_{j=1} \left[ \mathbf{r}_j(t) - \mathbf{r}_j(0) \right].
	\end{equation}
	We used cut-off value $r^{nn}_c=1.3$ for the first nearest neighbours $N_{nn}$. After identifying the $N_{nn}$ particles at the initial time (time origin) we track these particles dynamics at later time t. Using these displacement we compute $Q^{CR}(t)$, $\chi^{CR}_4(t)$, and MSD respectively.

\vskip +0.1in
\noindent{\bf \large Hessian Matrix ($\mathcal{H}$): } The Hessian matrix is defined as the double derivative of the total potential energy, $U(\{ \mathbf{r}_i\})$, as
	\begin{equation}
	\mathcal{H}_{ij}^{\alpha\beta} = \frac{\partial^2 U(\{ \mathbf{r}_i\})}{\partial r_i^\alpha \partial r_i^\beta},
	\end{equation}
	with $i,j$ being the particle index and $\alpha, \beta$ being the dimensionality index.

\vskip +0.1in
\noindent{\bf \large Data Availability: } All the data produced in this work are presented in the manuscript and its Supplementary information file. Source data for the figures are available in a public repository on Figshare at \href{https://doi.org/10.6084/m9.figshare.29234687}{https://doi.org/10.6084/m9.figshare.29234687} (ref. \cite{Deydatacode2025}). Raw Molecular Dynamics data will be available from the authors upon request.

\vskip +0.1in
\noindent{\bf \large Code Availability: } All the codes used in this work to produce the data are available in a public repository on Figshare at \href{https://doi.org/10.6084/m9.figshare.29234687}{https://doi.org/10.6084/m9.figshare.29234687} (ref. \cite{Deydatacode2025}).


\vskip +0.1in
\noindent{\bf \large Acknowledgments: }
We thank Chandan Dasgupta,  Gilles Tarjus,  Giulio Biroli,  Ludovic Berthier and Sumilan Banerjee for many useful discussions.  We specially acknowledge Sumilan Banerjee for bringing to our notice the existing studies on non-linear phonon-dispersion relation in fractal solids.  We acknowledge funding by intramural funds at TIFR Hyderabad from the Department of Atomic Energy (DAE) under Project Identification No. RTI 4007.  SK acknowledges the Swarna Jayanti Fellowship grants DST/SJF/PSA01/2018-19 and SB/SFJ/2019-20/05 from the Science and Engineering Research Board (SERB) and Department of Science and Technology (DST). Most of the computations are done using the HPC clusters procured using Swarna Jayanti Fellowship grants DST/SJF/PSA01/2018-19, SB/SFJ/2019-20/05 and Core Research Grant CRG/2019/005373.
SK also acknowledges research support from MATRICES Grant MTR/2023/000079 from SERB. 

\vskip +0.1in
\noindent{\bf \large Author Contributions:}
SK conceived the project.  SK supervised the project.  SD and AB performed research and simulations.  SD, AB, and SK designed analysis methods.  SD and AB performed all the analyses.  SD, AB, and SK wrote the paper jointly.  SD and AB contributed in this work equally.

\vskip +0.1in
\noindent{\bf \large Competing interests:}
All the authors declare no competing interests.

\end{document}


\newcommand{\rev}[1]{\textcolor{black}{{#1}}}
\renewcommand{\figurename}{\textbf{Supplementary Fig.}}

\title{\rev{Supplementary Information : Enhanced Long Wavelength Mermin-Wagner-Hohenberg Fluctuations in Active Crystals and Glasses}}
\author{Subhodeep Dey}
\thanks{equal contributions}
\affiliation{Tata Institute of Fundamental Research Hyderabad, 36/P, Gopanpally Village, Serilingampally Mandal, Ranga Reddy District,
Hyderabad, Telangana 500046, India}
\author{Antik Bhattacharya}
\thanks{equal contributions}
\affiliation{Tata Institute of Fundamental Research Hyderabad, 36/P, Gopanpally Village, Serilingampally Mandal, Ranga Reddy District,
Hyderabad, Telangana 500046, India}
\author{Smarajit Karmakar}\email{smarajit@tifrh.res.in}
\affiliation{Tata Institute of Fundamental Research Hyderabad, 36/P, Gopanpally Village, Serilingampally Mandal, Ranga Reddy District,
Hyderabad, Telangana 500046, India}

\maketitle

\section{Models and Methods}

In this study, we have used a Binary Mixture of particles, which interact via well-known Lennard-Jones (LJ) potential. This potential has been tuned such that the $2^{nd}$ derivative of the potential is smoothed up to the cut-off distance $r_c$,
\begin{equation}
 \phi(r) = \begin{cases}
 4\epsilon_{\alpha\beta}\left[\left(\frac{\sigma_{\alpha\beta}}{r}\right)^{12} -\left(\frac{\sigma_{\alpha\beta}}{r}\right)^{6} + c_0 + c_2r^2 \right] &, r < r_c \\
 0 &, r \geq r_c
 \end{cases}
\end{equation}
Here, $\{\alpha,\beta\}$ refers to the type of the particles, which can be A (large) or B (small) type. We studied two types of glass-forming liquids; one is a 2D modified Kob-Andersen (2dmKA), where the A:B number ratio is $65:35$, and the other one is a 2D Kob-Andersen model (2dKA), where the A:B number ratio is $80:20$ \cite{Tah2018}. The interaction strengths are $\epsilon_{AA}=1.0$, $\epsilon_{AB}=1.5$, $\epsilon_{BB}=0.5$ and the diameter of the particles are $\sigma_{AA}=1.0$, $\sigma_{AB}=0.8$, $\sigma_{BB}=0.88$. The interaction cut-off is $r_c=2.5\sigma_{AB}$. Here, the reduced unit of length, energy, and time is given by $\sigma_{AA}$, $\epsilon_{AA}$ and $\sqrt{\frac{\sigma^2_{AA}}{\epsilon_{AA}}}$ respectively. \rev{For the 3dKA and 4dKA system is the Kob-Andersen model with A:B=80:20 in 3 and 4 dimensions respectively}. For all cases, the system's particle density ($\rho$) is fixed at $1.2$, and the integration step ($\delta t$) is set at $0.005$. For polycrystal simulations, we used only A-type particles at a density of $1.2$, and for the integration step here, we also kept the same $\delta t = 0.005$. \rev{Similarly, 3d-polycrystal is same as 2d-polycrystal in 3 dimension.}

\section{Activity in the system: Run and Tumble particle (RTP) model}
To introduce the activity in the system, we have given additional active force, $f_0$, to a set of selected (tagged) active particles and their mutual particle-particle interaction coming from the inter-particle potential. Here, we have selected a fraction ($c$) of active particles in the system. We assign each active particle an active direction $k_{\alpha}$ for $[\alpha \in x,y]$ randomly in a manner that the total active force applied to all the active particles sums up to zero, which is done to maintain the momentum conservation in the system. The persistent time $\tau_p$ is the time until the active force on each particle acts in the same direction, and after $\tau_p$, the directions are randomly reshuffled. The active force on the $i^{th}$ particle in 2D can be written as,
\begin{equation}
\mathbf{F}^A_i = f_0 (k^i_x \hat{x}+k^i_y \hat{y}),
\end{equation}
where the active direction of the $i^{th}$ particle is $k^i_{\alpha}$ for $[\alpha \in x,y]$ is chosen from $\pm1$. The number of total active particles is taken to be an even number to maintain the total active momentum to be zero along all directions independently, i.e., $\sum_{\alpha,i} k^i_{\alpha}=0$. The basis set in 3d and 4d would be $[\alpha \in x,y,z]$ and $[\alpha \in x,y,z,w]$ respectively. In this RTP active model, there are three tuning parameters $c$, $f_0$, and $\tau_p$. We are going to study the effect of different activities on different glassy and polycrystalline systems. To start with, we first fixed $\tau_p = 1.0$ for all the cases, then we changed the $f_0$ and $c$ separately. For 2dmKA, we have varied $f_0$ $\in$ [0.0-2.0], keeping $c$ fixed at $0.1$. To study the effect of activity in the solid regime for disorder and polycrystalline systems, we have varied the activity in the energy-minimized system at temperature T=0.01.

\section{Thermostat}
In this study, we have used the three-chain Nos\'e-Hoover thermostat \cite{Allen} to get the required temperature. This thermostat is known to give a true canonical ensemble in an equilibrium system, which is also suitable for a canonical ensemble in an out-of-equilibrium system. Here, we have taken the coupling relaxation time ($\tau_T$) of the thermostat $10$ times the integration time step ($\delta t$).

\section{Two-point correlation Function, $Q(t)$}
To compute two-point density-density correlation, we have considered a simpler form of overlap correlation function $Q(t)$ defined as
\begin{equation}
 Q(t) = \frac{1}{N} \sum^N_{i=1} w(|\mathbf{r}_i(t)-\mathbf{r}_i(0)|),
\end{equation}
where $w(x)$ is a window function, which is $1$ for $x<a$ and $0$ otherwise. $\mathbf{r}_i(t)$ is the position of the $i^{th}$ at time $t$. Here, parameter `$a$' is chosen to remove the vibration due to the caging effect. One can choose the value of 'a' from the plateau of a `mean-square displacement' (MSD) in the supercooled temperature regime. For all cases, the value of $a$ is set to $0.3$. From this two-point correlation function, we define a relaxation time $\tau_{\alpha}$, as $\left<Q(t=\tau_{\alpha})\right>=1/e$, where $\left<...\right>$ is ensemble average and $e$ is the base of the natural logarithm. The system is equilibrated for $150\tau_{\alpha}$, and then we ran the simulations for another $150\tau_{\alpha}$ for measuring various dynamical and thermodynamic quantities. We chose $\tau_{\alpha} \sim 2000$. For system size $N \leq 10000$, we have averaged our data over $32$ statistically independent ensemble runs, and for system size $N > 10000$, we have taken $6$ ensembles for averaging.

\section{Cage-relative displacement}
To separate the influence of the collective behavior affecting the system dynamics due to the glassy relaxation process and not from the long wavelength phonon fluctuations, we have computed the cage-relative (CR) displacement \cite{Mazoyer2009, Vivek2017, Illing2017} of the individual particle $i$, such as
\begin{equation}
 \mathbf{r}_{i,CR}(t) = \left[ \mathbf{r}_i(t) - (\mathbf{r}_{i,nn}(t) - \mathbf{r}_{i,nn}(0) ) \right],
\end{equation}
where $\mathbf{r}_{i,nn}(t)$ is the position of the center of mass of $N_{nn}$ nearest neighbours of $i^{th}$ particles. Again, it is defined as
\begin{equation}
 \mathbf{r}_{i,nn}(t) = \frac{1}{N_{nn}} \sum^{N_{nn}}_{j=1} \left[ \mathbf{r}_j(t) - \mathbf{r}_j(0) \right].
\end{equation}
Here, we have used cut-off value $r^{nn}_c=1.3$ for the first nearest neighbours $N_{nn}$. After identifying the $N_{nn}$ particles at the initial time or at each time origin, we track these particles' ($N_{nn}$) dynamics at the later time for the computation of the cage relative correlation functions. We use this cage-relative displacement to calculate $Q(t)$, and MSD, respectively.

\section{choice of temperature for different activity}
\begin{figure*}[!htpb]
\centering
\includegraphics[width=0.82\textwidth]{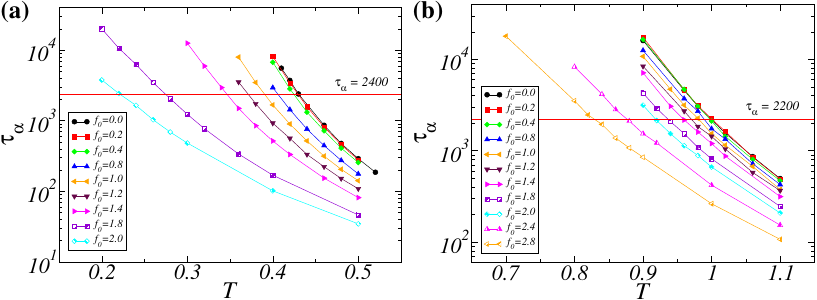}
\caption{(a) $\tau_{\alpha}$ vs T for different $f_0$ and fixed c=0.1 and $\tau_p=1.0$, for 2dmKA model. Similarly, (b) for the 2dKA model. From these plots, we can estimate the temperature for the corresponding activity, which can give similar relaxation time ($\tau_{\alpha}$) for different activities for system size N=1000.}
\label{fig:tauAlphavsT_f0_2dmKA_2dKA_merge}
\end{figure*}
As activity in each model (2dmKA and 2dKA) can be tuned either by tuning the fraction of active particles ($c$) or the active force ($f_0$) or the persistent time ($\tau_p$), it is important to identify a correct measure to compare these systems across the parameter space. For this, we choose relaxation time $\tau_{\alpha}$ as the system's characteristics for comparison across various values of $c$, $f_0$, and $\tau_p$. For a given set of $c$, $f_0$ and $\tau_p$, if we vary $T$, then the obtained relaxation time is found to obey the well-known `Vogel-Fulcher-Tammann' (VFT) relation, defined as
\begin{equation}
 \tau_{\alpha} = \tau_0 \exp[A/T-T_0].
\end{equation}

We then take an intermediate system size and fix the $\tau_{\alpha}$ for different activities at the suitable temperature using the VFT relation. For 2dmKA, we have taken the relaxation time $\tau_{\alpha} \simeq 2400$, which happens at $T = 0.43$ of a passive system with $N = 1000$. We then fix these temperatures for all other system sizes to get the system size effect for a fixed temperature with corresponding activity in the supercooled liquid regime. For 2dmKA model system, we firstly change the $f_0$ from $0.0$ to $2.0$ by fixing $c = 0.1$ and $\tau_p = 1.0$. Similarly, for 2dKA, we have taken the relaxation time $\tau_{\alpha} \simeq 2200$ at $T = 1.00$ of a passive system with $N = 1000$ and fixed it for all other activities and system sizes. We changed the values of $f_0$ from $0.0$ to $2.8$ by fixing $c = 0.1$ with $\tau_p = 1.0$. From Supplementary Fig. \ref{fig:tauAlphavsT_f0_2dmKA_2dKA_merge}, we have used the temperature for the respective activity parameter and fixed it for the rest of the study of glassy dynamics of supercooled liquid in 2d. Again, in Supplementary Fig. \ref{fig:relax_N_1e3_f0_2dmKA_2dKA_merge}, we have shown that the choice of temperature corresponding to different activity is giving similar relaxation time for the system size $N = 1000$.

\begin{figure*}[!htpb]
\centering
\includegraphics[width=0.82\textwidth]{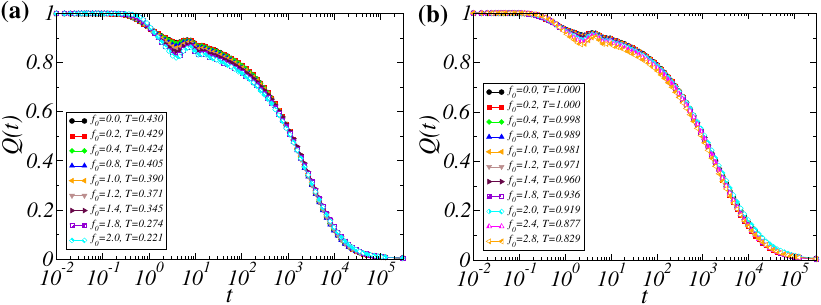}
	\caption{(a) Two-point correlation vs for different $f_0$ at fixed c=0.1, $\tau_p=1.0$ gives similar relaxation time for 2dmKA system. Similarly, (b) relaxation dynamics for different activity $f_0$ for 2dKA system.}
\label{fig:relax_N_1e3_f0_2dmKA_2dKA_merge}
\end{figure*}

\section{Mean square displacement from the dynamical matrix}
Taking the following Hamiltonian to describe the system
\begin{equation}
\mathcal{H}=\sum_{i=1}^{N}\frac{{\mathbf{p}_i}^2}{2m_i}+\frac{1}{2}\sum_{i \neq j} U(\mathbf{r}_i, \mathbf{r}_j),
\end{equation}
if we assume that the harmonic approximation will be valid at low temperatures; position of the particles around an inherent structure(IS) can be written as:
\begin{equation}
	\mathbf{r}_i = \mathbf{r}_{i,0} + \mathbf{u}_i
\end{equation}
where, $\mathbf{r}_i$, $\mathbf{r}_{i,0}$ and $\mathbf{u}_i$ represents the molecular dynamics trajectory, IS trajectory and small displacement from IS, respectively.\\
The Hamiltonian can be rewritten as:
\begin{equation}
	\mathcal{H} \simeq \sum_{i=1}^{N}\frac{{\mathbf{p}_i}^2}{2m_i} +\frac{1}{2}H^{\alpha \beta}_{ij}u_i^{\alpha}u_j^{\beta} + V(\{\mathbf{r}_{i,0}\})
\end{equation}
where, $V(\{\mathbf{r}_{i,0}\})=\frac{1}{2}\sum_{i \neq j} U(\mathbf{r}_i, \mathbf{r}_j)$. Lets take $m_i=1$.
Now the equation of motion becomes:
\begin{equation}
	\ddot{\mathbf{u}}_i = -\sum_{j \beta} H^{\alpha \beta}_{ij}u^{\beta}_j
\end{equation}
where, $H^{\alpha \beta}_{ij}u^{\beta}_j = \left[ \frac{\partial^2 V}{\partial r^\alpha_i \partial r^\beta_j}\right]_0$

Redefining the variables, $\{u^{\alpha}_i\} \rightarrow \{ q_a\}$ and $\{\dot{u^{\alpha}_i}\} \rightarrow \{ \dot{q_a}\}$; where $i=1, 2,...,N$, $\alpha=1,2$ and $a = 1,2,...,2N$. Here, $a=(i-1)d+\alpha$ where $d$ is the dimension. So, $u^{\alpha}_i = q_a = q_{(i-1)d+\alpha}$. Now the Hamiltonian becomes,
\begin{equation}
	\mathcal{H} = V_0 + \frac{1}{2}\sum_i \dot{q_a}^2 + \frac{1}{2}\sum_{a,b} H_{ab} q_a q_b
\end{equation} 
And the equation of motions transforms as,
\begin{equation}
	\ddot{q_a} = -\sum_b H_{ab} q_b
\end{equation}
Now by performing an orthogonal transformation on $\mathbf{\mathcal{H}}$ we get,
\begin{equation}
\begin{split}
	\mathbf{\xi} &= \mathbf{S}. \mathbf{q}\\
	\mathbf{S}. \mathbf{\ddot{q}} &= -\mathbf{S}.\mathbf{\mathcal{H}} . \mathbf{S}^{-1}.\mathbf{S}.\mathbf{q}
\end{split}
\end{equation}
So that, $\mathbf{D} = \mathbf{S}.\mathbf{\mathcal{H}}.\mathbf{S}^{-1}$ is diagonal, $D_{ab} = \lambda_a. \delta_{ab}$\\
To find $\mathbf{S}$ and $\{\lambda_a\}$ diagonalize $\mathbf{\mathcal{H}}$, {\it i.e}
\begin{equation}
	\sum_b \mathcal{H}_{ab} o^{(n)}_b = \lambda_n o^{(n)}_a
\end{equation}
Define $S_{ab} = o^{(n)}_a$ is orthogonal. The equations of motion for normal modes,
\begin{equation}
\begin{split}
	\ddot{\mathbf{\eta}} &= -\mathbf{D}.\mathbf{\eta}\\
\text{so,    }	\ddot{\mathbf{\eta_a}} &= -\lambda_a.\mathbf{\eta_a}\\
\end{split}
\end{equation}
Solving the above equation we get,
\begin{equation}
	\eta_a(t) = \eta_a(0)cos(\omega_a t) + \dot{\eta_a}(0)\frac{sin(\omega_a t)}{\omega_a}
\end{equation}
Now,
\begin{equation}
\begin{split}
\mathbf{\eta}(t) &= \mathbf{S}.\mathbf{q}(t)\\
\mathbf{q}(t) &= \mathbf{S}^{-1}.\mathbf{\eta}(t)\\
q_a(t) &= \sum_b (S^{-1})_{ab}.\eta_b(t)
\end{split}
\end{equation}
or, 
\begin{equation}
	q_a(t) = \sum_{b} o^b_a .\eta_b(t)
\end{equation}
Then the actual molecular dynamics trajectories can be written as:
\begin{equation}
	r_i^{\alpha}(t) = r_i^{\alpha}(0) + \sum_{b}o^{(b)}_{(i-1)d+\alpha}\left[\eta_b cos(\omega_b t) + \dot{\eta_b}\frac{sin(\omega_b t)}{\omega_b} \right]
\end{equation} 
In the above summation, zero modes are excluded.

Now we will calculate the mean squared displacement using the above formalism,
\begin{widetext}
\begin{equation}
\begin{split}
	(r_i^{\alpha}(t)-r_i^{\alpha}(0))^2 &= \sum_{\alpha, a, b} o^{(a)}_{(i-1)d+\alpha}o^{(b)}_{(i-1)d+\alpha}
	[\eta_a \eta_b(cos(\omega_a t)-1) (cos(\omega_b t)-1)+\\ &2\dot{\eta_a}\dot{\eta_b}\frac{sin(\omega_a t)}{\omega_a}
	(cos(\omega_b t) -1) +\dot{\eta_a}\dot{\eta_b}\frac{sin(\omega_a t)}{\omega_a}\frac{sin(\omega_b t)}{\omega_b}]
\end{split}
\end{equation}
\end{widetext}
So, after taking time origin average and ensemble average it becomes:
\begin{widetext}
\begin{equation}
\begin{split}
	<(r_i^{\alpha}(t)-r_i^{\alpha}(0))^2> &= \sum_{\alpha, a, b} o^{(a)}_{(i-1)d+\alpha}o^{(b)}_{(i-1)d+\alpha}
	[<\eta_a \eta_b>(cos(\omega_a t)-1) (cos(\omega_b t)-1)+\\ &2<\dot{\eta_a}\dot{\eta_b}>\frac{sin(\omega_a t)}{\omega_a}
	(cos(\omega_b t) -1) +<\dot{\eta_a}\dot{\eta_b}>\frac{sin(\omega_a t)}{\omega_a}\frac{sin(\omega_b t)}{\omega_b}]
\end{split}
\end{equation}
\end{widetext}
$<\eta_a \eta_b>=0$ for $a \neq b$ and for $a=b$
\begin{equation*}
	<\eta_a \eta_b> = <\eta_a^2> = \frac{\int d\eta_a e^{-\frac{w_a^2 \eta_a^2}{2K_BT}}\eta_a^2}{\int d \eta_a e^{-\frac{w_a^2 \eta_a^2}{2K_BT}}}=\frac{K_B T}{\omega_a^2}
\end{equation*}
Similarly, $<\dot{\eta_a}\eta_b>=0$ and $<\dot{\eta_a}\dot{\eta_b}>=K_B T \delta_{ab}$\\
Now,
\begin{widetext}
\begin{equation}
\begin{split}
	<(r_i^{\alpha}(t)-r_i^{\alpha}(0))^2> = \sum_{\alpha, a}\left(o^{(a)}_{(i-1)d+\alpha}\right)^2 \left[K_B T \frac{(cos(\omega_a t)-1)^2}{\omega_a^2}+K_B T \frac{sin^2(\omega_a t)}{\omega_a^2}\right]
\end{split}
\end{equation}
\end{widetext}
So, Mean square displacement
\begin{equation}
 <\Delta r(t)^2> = \frac{K_B T}{N} \left[ \sideset{}{'}\sum_{a,i} (\mathbf{P}^a_i)^2 \frac{sin^2(\omega_a t/2)}{(\omega_a /2)^2}\right].
 \label{Eq:MSD}
\end{equation}
Where $a$ is for all degrees of freedom ($d\cdot N$), and i is for all particles in the system N. $\omega_a$ is the $a^{th}$ eigen frequency and $\mathbf{P}^a_i$ is the $i^{th}$ eigen vector corresponding to $\omega_a$. The prime summation denotes that it is for non-zero eigenmodes only. 

\section{Analysis Protocol of Dynamical Matrix}
Using the trajectories of particles, we calculate the displacement correlation matrix $C_{ij}$, which captures both local and long-range correlations in the particle motion. $C_{ij}$ is defined as:
\begin{equation}
	C_{ij}=\left< \left(\mathbf{r}_i(t)-\left< \mathbf{r}_i\right>\right) \cdot \left(\mathbf{r}_j(t)-\left< \mathbf{r}_j\right>\right)\right>
\end{equation}  
where $\left< \dots \right>$ represents the time average.

The relationship between the dynamical matrix and the actual Hessian of the system is given by $C_{ij}^{\alpha \beta} = \frac{1}{T} (\mathcal{H}_{ij}^{\alpha \beta})^{-1}$. Over time, $C_{ij}^{\alpha \beta}$ is expected to converge to $\mathcal{H}_{ij}^{\alpha \beta}$ as $t \rightarrow \infty$. However, practical constraints arise due to the limited computational runs and frames available for matrix generation. The eigenvalue distribution of the actual Hessian typically exhibits only $d$ zero eigenmodes, where $d$ represents the system's dimension. If a few frames are used to compute $C{ij}^{\alpha \beta}$, the resulting distribution may show more than $d$ zero eigenmodes. Nevertheless, increasing the number of frames tends to reduce this discrepancy. To validate this observation, we analyzed a passive system and computed its Hessian, which enabled us to derive the density of states (DoS). As we progressively increased the number of frames, the DoS obtained from the correlation matrix increasingly converged to that obtained from the Hessian.

\begin{figure}[!htpb]
	\centering
	\resizebox{90mm}{70mm}{\includegraphics{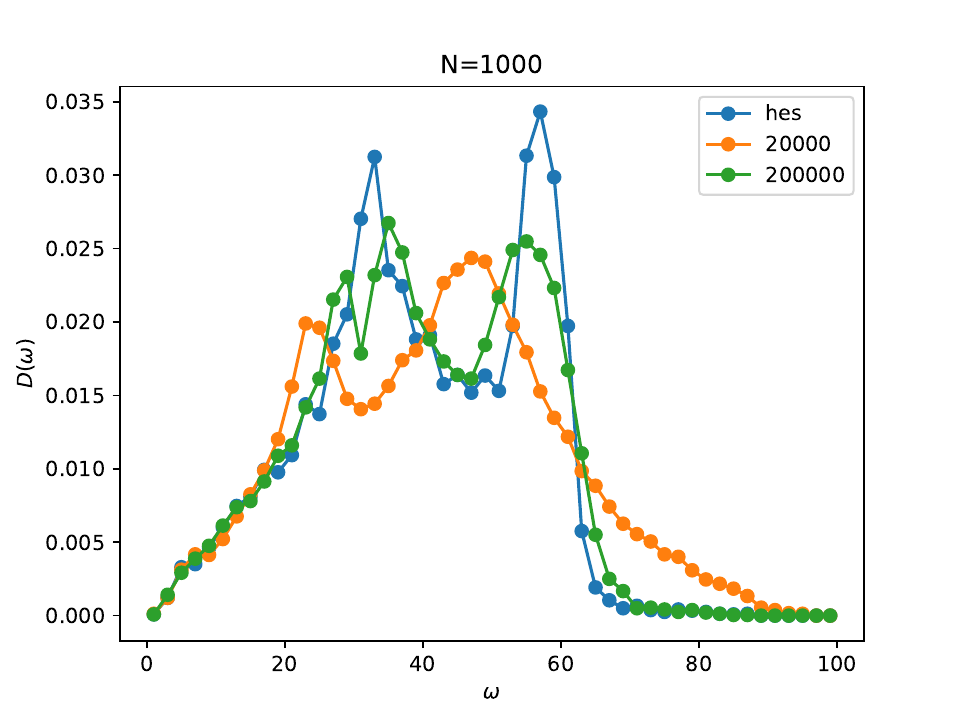}}
	\caption{$D(\omega)$ vs. $\omega$ for different number of frames from disp-disp correlation matrix compared with the DoS from hessian, poly-crystal system size $N=1000$}
	\label{fig:uu_DOS_passive_N1000}
\end{figure}
However, the number of frames required for convergence depends entirely on the system size. As the system size increases, the necessary number of frames also increases rapidly. To quantify this, $R = N/T$ emerges as a crucial parameter, where $T$ represents the number of independent time frames used for constructing the covariance matrix, and $N$ denotes the number of degrees of freedom in the system. A $R < 1$ value ensures that the covariance matrix is constructed from independent measurements, while $R \rightarrow 0$ corresponds to the limit of perfect statistics. As $R$ increases, the noise in the dynamical matrix also increases, leading to systematic errors in the density of states (DoS). Random matrix theory suggests that the eigenvalue distribution should converge linearly to its limiting values for disordered systems when $R=0$.

{\bf Correction of DoS:} As illustrated in Supplementary Fig-\ref{fig:omega_correction}, eigenfrequencies $\omega_n$ exhibits a linear dependence on $R$. Thus, we analysed higher $R$ values and extrapolated the data to $R \to 0$ to obtain the corrected eigenfrequencies. This correction significantly improved the convergence of the new DoS. A comparison between the corrected DoS and the DoS obtained from the Hessian is presented in the main text.
\begin{figure}[!htpb]
	\centering
	\resizebox{90mm}{70mm}{\includegraphics{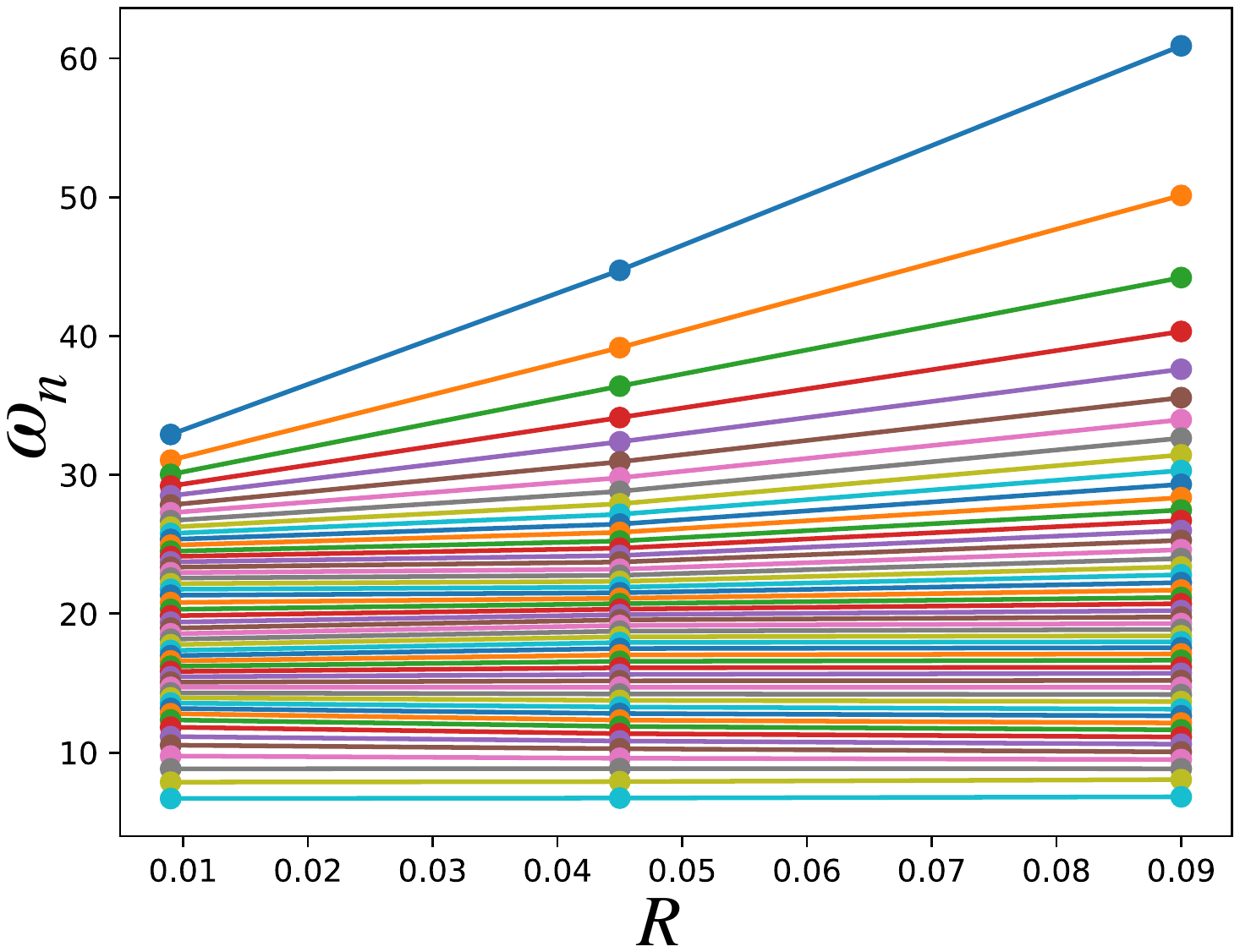}}
	\caption{eigen modes vs. R for system size $N=1000$ poly-crystal}
	\label{fig:omega_correction}
\end{figure}

{\bf Force-force Correction Matrix:} Similar to the displacement-displacement correlation matrix, we performed the same analysis for the force-force correlation matrix to get the DoS. $F_{ij}^{\alpha \beta}$ is defined as
\begin{equation}
F_{ij}^{\alpha \beta}=f_i^{\alpha}f_j^{\beta}
\end{equation}  
\begin{figure}[!htpb]
	\centering
	\resizebox{90mm}{70mm}{\includegraphics{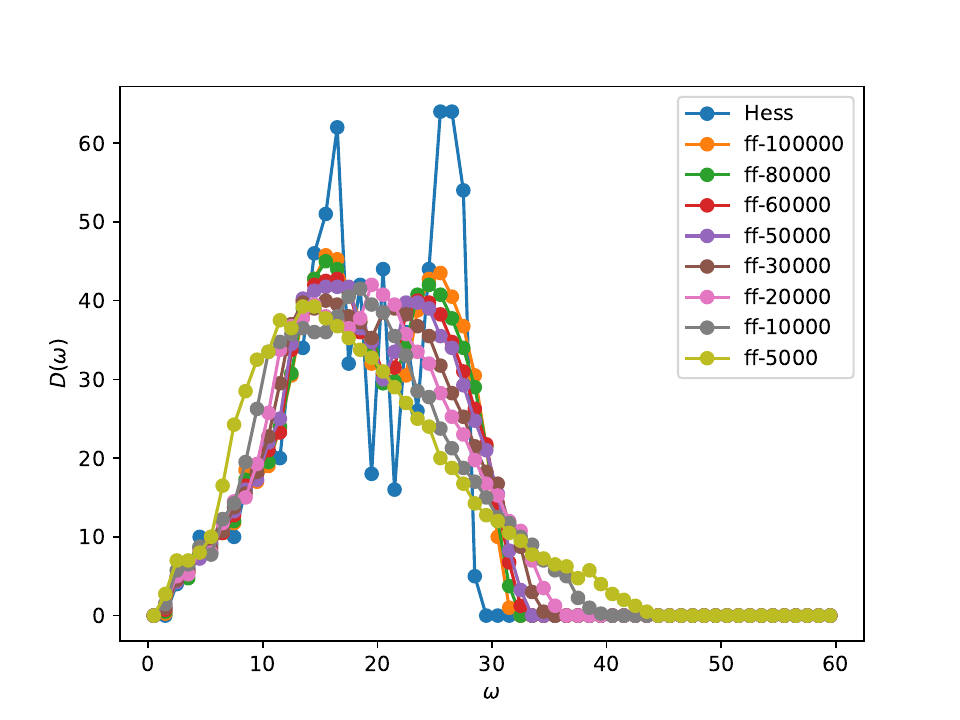}}
	\caption{$D(\omega)$ vs. $\omega$ for different numbers of frames from the ff-correlation matrix compared with the DoS from Hessian, system size $N=400$ for crystal.}
	\label{fig:ff_DOS_N400}
\end{figure}
\begin{figure}[!htpb]
	\centering
	\resizebox{90mm}{70mm}{\includegraphics{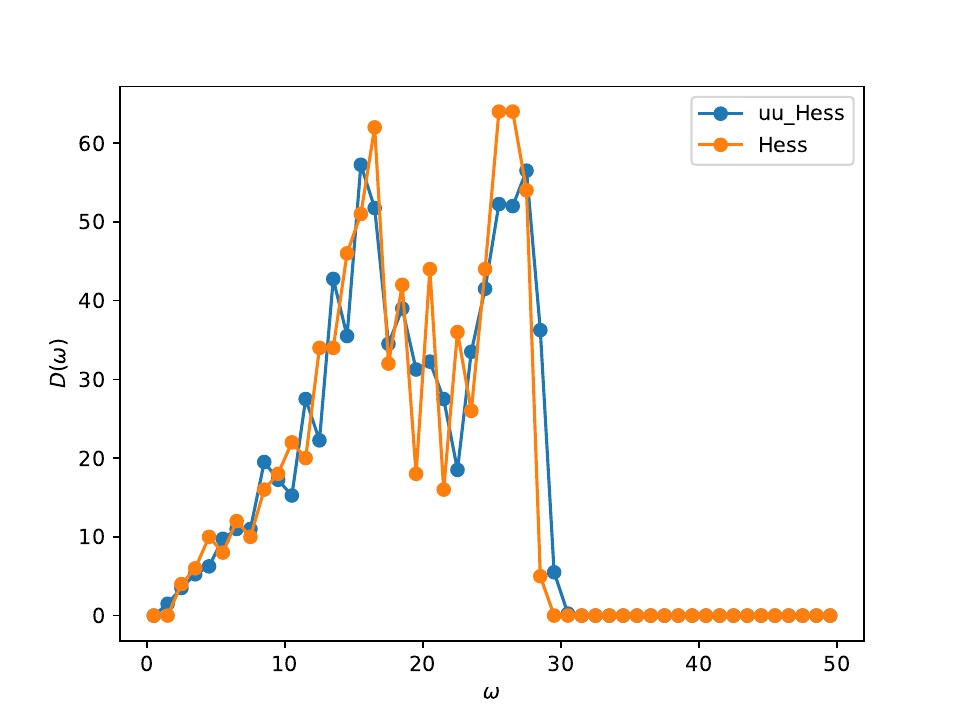}}
	\caption{Corrected DoS from ff-correlation matrix for system size $N=400$ for crystal}
	\label{fig:ff_corrected_DOS_N400}
\end{figure}
For the passive system, the convergence was pretty good (see Supplementary Fig-\ref{fig:ff_corrected_DOS_N400}).

\begin{figure}[!htpb]
	\centering
	\resizebox{85mm}{65mm}{\includegraphics{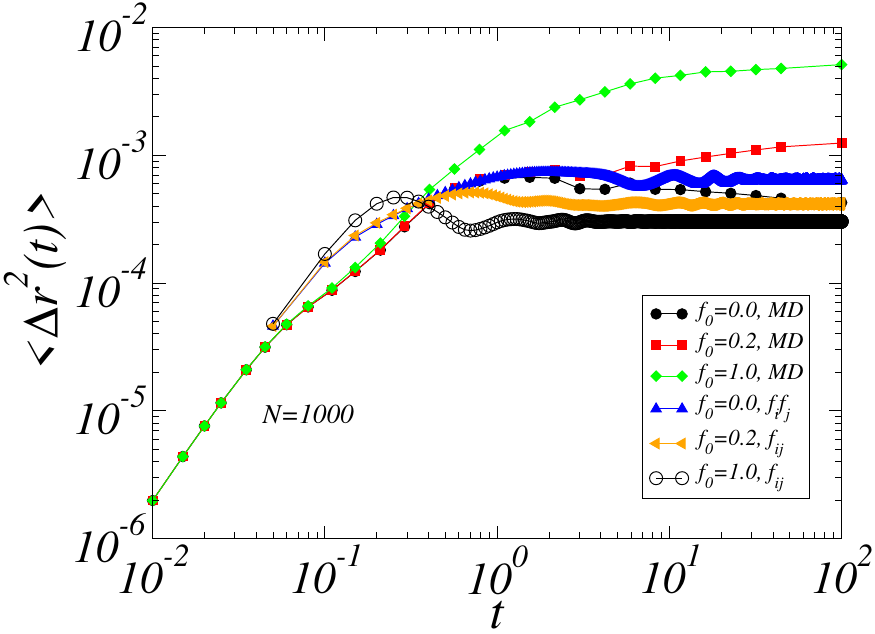}}
	\caption{MSD from molecular dynamics compared with MSD from ff-correlation matrix generated with $100000$ frames, system size $N=1000$}
	\label{fig:MSD_t_avgff_f0_N_1e3}
\end{figure}

Using the eigenvectors and eigenvalues from the dynamical matrix, one can calculate the mean square displacement (MSD) via Eqn.\ref{Eq:MSD}. We employed this approach to verify if the density of states (DoS) from the dynamical matrix provides accurate information about the system. We observed that the MSD from both force-force and displacement-displacement correlation matrices for passive systems matched the Molecular Dynamics MSD. However, for active systems, only the MSD from the displacement-displacement correlation matrix showed good agreement (refer to the main text). In contrast, the force-force correlation MSD exhibited an opposite trend (refer to Supplementary Fig-\ref{fig:MSD_t_avgff_f0_N_1e3}). Therefore, for all further analyses, we utilized the displacement-displacement correlation.

{\bf Dispersion relation analysis:} 
For each eigenmode obtained from the covariance matrix, the Fourier transform of its longitudinal and transverse components yields two spectral functions, denoted as $f_L$ and $f_T$, respectively
\begin{equation}
	f_T(q,\omega)=\left< \left| \sum_{j=1}^n\widehat{\bf{q}} \times \bf{e}_{\omega,j}e^{i\bf{q}\cdot\bf{r}_j}\right|^2\right>
\end{equation}
\begin{equation}
f_L(q,\omega)=\left< \left| \sum_{j=1}^n\widehat{\bf{q}} \cdot \bf{e}_{\omega,j}e^{i\bf{q}\cdot\bf{r}_j}\right|^2\right>
\end{equation}
where $\bf{e}_{\omega,j}$ is the polarization vector on the particle $j$ in mode $\omega$, $\bf{q}$ is the wave vector.

\begin{figure}[!htpb]
	\centering
	\resizebox{85mm}{65mm}{\includegraphics{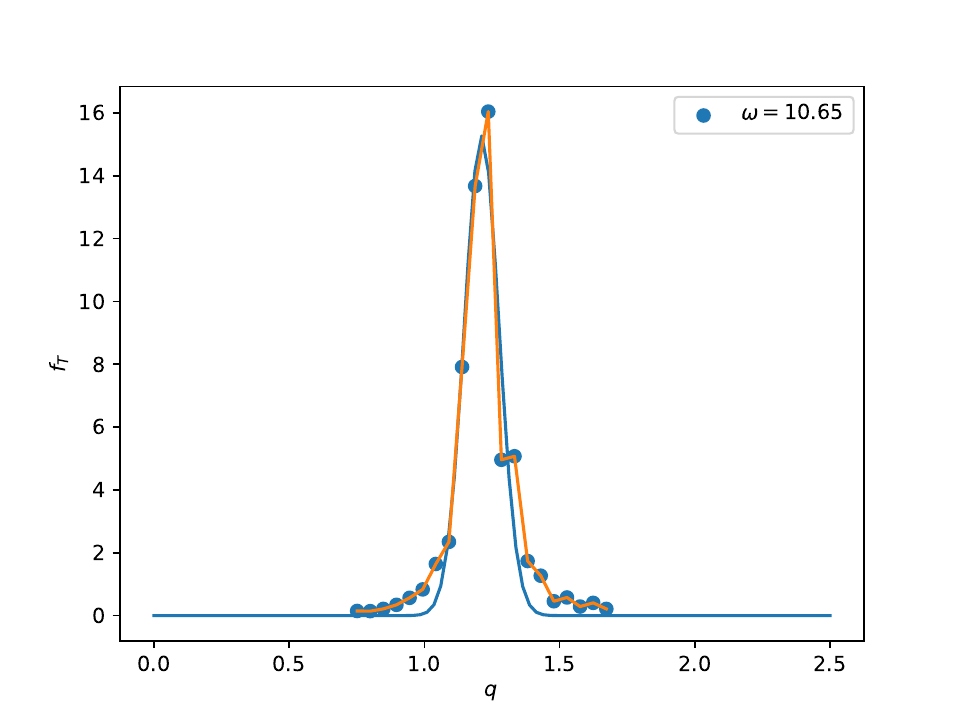}}
	\caption{$f_T$ vs $q$ for $\omega=10.65$, $N=4000$ poly-crystal and $f_0=1.0$. ($q_{max}=1.21$)}
	\label{fig:fT_fit}
\end{figure}

\begin{figure}[!htpb]
	\centering
	\resizebox{85mm}{65mm}{\includegraphics{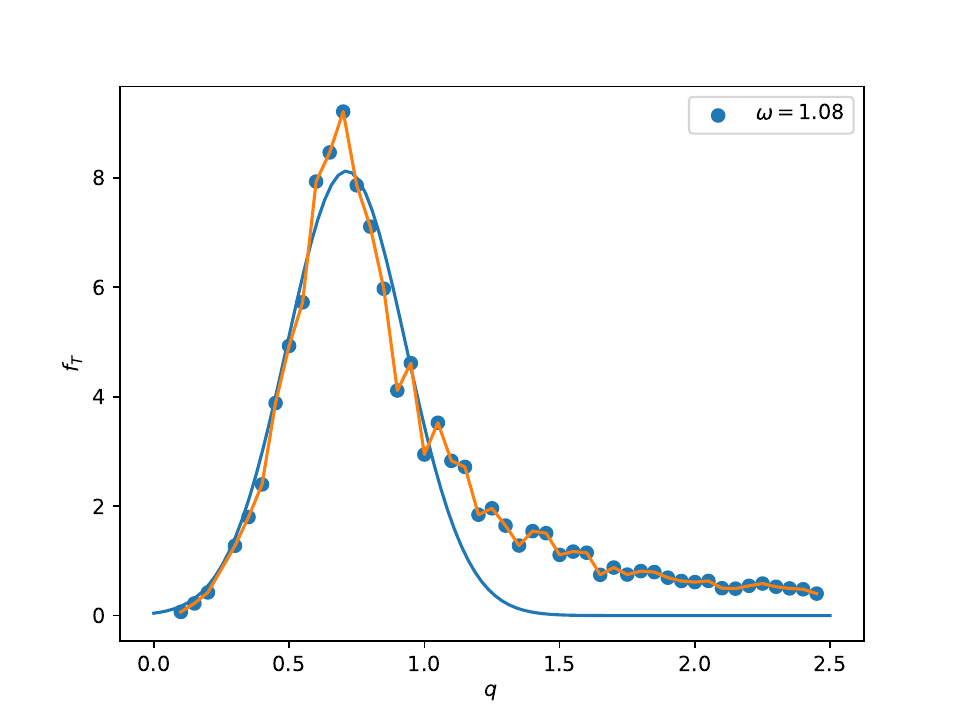}}
	\caption{$f_T$ vs $q$ for $\omega=1.08$, $N=4000$ glass and $f_0=1.0$. ($q_{max}=0.712$)}
	\label{fig:fT_fit_glass}
\end{figure}

Next, using the above formula, we binned $f_L$ and $f_T$ in the $q$-$\omega$ plane (refer to the heat map in the main text). Then, for each $\omega$-bin, we fitted a Gaussian function in the vicinity of the peak of the $f_L$ vs $q$ and $f_T$ vs $q$ curves to obtain the corresponding $q_{\text{max}}$. This method allows us to determine the dispersion curve by extracting the peaks (refer to Supplementary Fig \ref{fig:fT_fit}-\ref{fig:fT_fit_glass}).

\section{Derivation of the Dynamical Matrix calculation}
Here we present the mathematical derivation of the inverse relation between the displacement-displacement covariant matrix $\mathcal{C}$ with the Hessian matrix $\mathcal{H}$.  We start by writing the potential energy as 
\begin{equation}
U = U_0 + \frac{1}{2}\braket{u|\mathcal{H}|u} + \mathcal{O}(u^3) + \cdots
\end{equation}
 within Harmonic approximation keeping only up to squared displacements $|u>$ in the expansion and $\mathcal{H}$ is the hessian or the dynamical matrix of the system. If $\ket{\psi_n}$ is the eigenvector of the Hessian matrix, such that 
\begin{equation}
 \mathcal{H}\ket{\psi_n} = \lambda_n \ket{\psi_n},
\end{equation}
then we can write any vector using eigenvectors as basis set as $\ket{u} = \sum_n C_n \ket{\psi_n}$ and then rewrite the energy function as 
\begin{equation}
U = U_0 + \frac{1}{2}\sum_{m,n}C_n C_m\braket{\psi_n|\mathcal{H}|\psi_m} = U_0 + \frac{1}{2}\sum_n C_n^2 \lambda_n,
\end{equation}
where $C_n$ are the amplitude of the $n^{th}$ eigenvector on that given displacement vector $\ket{u}$. Given this, one can then compute the averaged displacement-displacement covariant matrix within Harmonic approximation (with $\mathcal{Z_N}$ being the configurational partition function) as 
\begin{widetext}
\begin{align*}
 \left<\ket{u}\bra{u}\right> &= \left<\sum_{m,n} C_m C_n \ket{\psi_m}\bra{\psi_n}\right> 
 \propto \frac{1}{\mathcal{Z_N}} \int \mathcal{D}C_p\ exp(-\beta /2 \sum_p C^2_p\lambda_p) \sum_{m,n} C_m C_n \ket{\psi_m}\bra{\psi_n}\\
 &=\sum_{m,n} \frac{\ket{\psi_m}\bra{\psi_n}}{\mathcal{Z_N}} \int \mathcal{D}C_p\ exp(-\beta /2 \sum_p C^2_p\lambda_p) C_m C_n \\
 &=\sum_{m,n} \ket{\psi_m}\bra{\psi_n} \left[\frac{\int dC_n \int dC_m exp(-\beta /2 (C^2_m\lambda_m + C^2_n\lambda_n)) \cdot C_n \cdot C_m}{\int dC_n \int dC_m exp(-\beta /2 (C^2_m\lambda_m + C^2_n\lambda_n))}\right] \\
 &=\sum_{m,n} \ket{\psi_m}\bra{\psi_n} \left[\frac{1}{\beta \lambda_n}\cdot \delta_{nm}\right] \\
 &= k_BT \sum_{m,n} \frac{\ket{\psi_m}\bra{\psi_n}}{\lambda_n} \\
 &=k_BT \mathcal{H}^{-1}.
\end{align*}
\end{widetext}
Thus, we obtain the following relation between the displacement-displacement covariant matrix $\mathcal{C}$ and the Hessian matrix $\mathcal{H}$ as
\begin{align}
 \mathcal{C} = \left<\ket{u}\bra{u}\right> = k_BT \cdot \mathcal{H}^{-1}.
\end{align}

\section{Additional Relaxation Time data}
In Supplementary Fig. \ref{fig:tauAlphavsT_CM_C_f0_2dmKA_2dKA_merge} shows the $\tau^{CR}_{\alpha}$ as a function of temperature for both 2dmKA and 2dKA system.
\begin{figure*}[!htpb]
\centering
\includegraphics[width=0.95\textwidth]{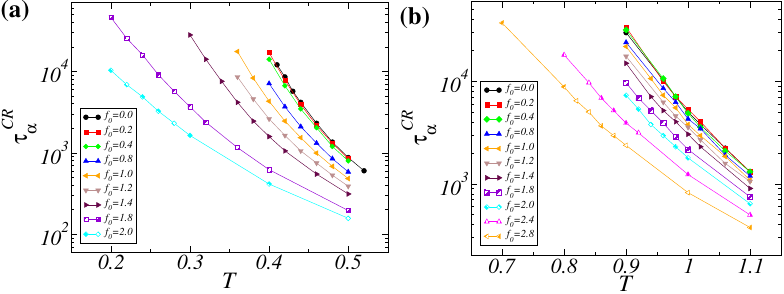}
\caption{(a) plot represents $\tau^{CR}_{\alpha}$ vs T for different $f_0$ and fixed c=0.1 and $\tau_p=1.0$, 2dmKA model (b) plots  represents $\tau^{CR}_{\alpha}$ vs T for different c and fixed $f_0=1.0$ and $\tau_p=1.0$, 2dKA model.}
\label{fig:tauAlphavsT_CM_C_f0_2dmKA_2dKA_merge}
\end{figure*}
We used these data and the measured diffusion constant to study the Stokes-Einstein breakdown in these systems. In Supplementary Fig. \ref{fig:DvstauAlpha_CM_0_f0_2dmKA_2dKA_merge}, we show the diffusivity $D$ as a function of $\tau_{\alpha}$ for all studied temperatures. We observe the breakdown of SE relation at a high-temperature regime due to long-wavelength fluctuation in a 2D system, which gets corrected once we remove the effect of long wavelength fluctuation via cage relative measurements or via Brownian dynamics simulations as in Supplementary Fig. \ref{fig:plot_merge/DvstauAlpha_D0_f0_2dmKA_merge} with increasing damping $D_0$ the phonon modes get suppressed, and one then recovers back the usual Stokes-Einstein relation at high temperature, thereby proving that anomalous SE breakdown at high temperatures in these active liquids is also due to long-wavelength phonon modes.

\begin{figure*}[!htpb]
\centering
\includegraphics[width=0.95\textwidth]{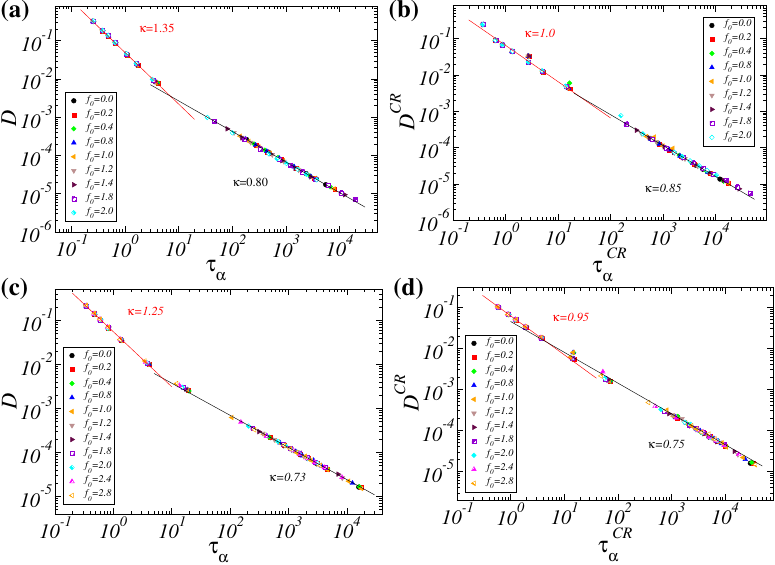}
\caption{D vs $\tau_{\alpha}$ shows D $\propto$ $\tau_{\alpha}^{-\kappa}$ in the supercooled regime for all active and passive system (a) \& (b) for 2dmKA model, (c) \& (d) for 2dKA model.}
\label{fig:DvstauAlpha_CM_0_f0_2dmKA_2dKA_merge}
\end{figure*}
\begin{figure*}[!htpb]
\centering
\includegraphics[width=0.90\textwidth]{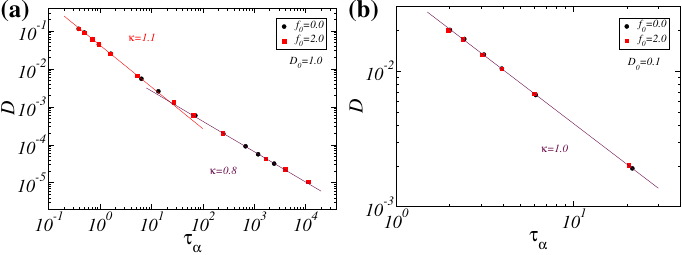}
\caption{D vs $\tau_{\alpha}$ shows D $\propto$ $\tau_{\alpha}^{-\kappa}$ in supercooled regime for all active and passive 2dmKA system with different damping (a) $D_0$=1.0 \& (b) $D_0$=0.1.}
\label{fig:plot_merge/DvstauAlpha_D0_f0_2dmKA_merge}
\end{figure*}

\begin{figure*}[!htpb]
\centering
\includegraphics[width=0.94\textwidth]{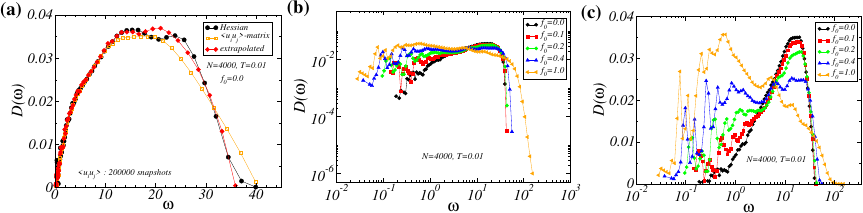}
\caption{(a) For passive systems of size N=4000, DoS is computed using the Hessian matrix (black circle), averaged displacement-displacement correlation matrix (orange square), and random matrix method (red diamond). (b \& c) DoS is computed using a displacement-displacement correlation matrix with increasing activity. Small frequency ($\omega$) peak in the DoS increases with increasing activity, which shows the enhancement of phonon-like modes. Also, in the small $\omega$ regime of DoS, there is jamming to unjamming transition due to increasing activity that can be observed. All plots are for the 2dmKA model.}
\label{fig:VDOS_log_f_000_f0_N_4e3_2dmKA_merge}
\end{figure*}

\begin{figure*}[!htpb]
\centering
\includegraphics[width=0.90\textwidth]{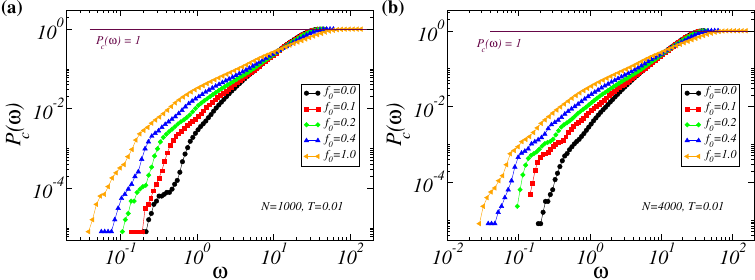}
\caption{The cumulative density of states $P_c(\omega)$ for the system size (a) N=1000 and (b) N=4000. Here, we have shown that the initial part of the effective VDoS increases rapidly with increasing activity. All plots are for the 2dmKA model.}
\label{fig:cumulative_log_f0_N_1e3_4e3_2dmKA_merge}
\end{figure*}

\section{Additional Analysis of VDoS}
In Supplementary Fig. \ref{fig:VDOS_log_f_000_f0_N_4e3_2dmKA_merge}, we show the vibrational Density of States (VDoS) of the 2dmKA model for all studied activities. A systematic increase in the weight of low-frequency modes in the system with increasing activity is curiously similar to the VDoS observed in jamming to unjamming transition in soft sphere assemblies, although in unjamming transition, one does not get an enhancement of phonon-like modes in the system. We believe that the enhancement of low-frequency modes in active systems is primarily due to the coupling of active forces with the shear modes of the systems. However, a detailed systematic analysis is needed to ascertain this angle of thought. We have a brief discussion on this aspect in the conclusion section of the main article.
In Supplementary Fig. \ref{fig:cumulative_log_f0_N_1e3_4e3_2dmKA_merge}, we show the cumulative distribution function, which seems to suggest that the initial power is close to $5$ for a passive system, which then systematically changes to smaller power with increasing activity, although one needs much better statistics to reliably estimate the power law at low frequencies, so we do not want to discuss possible connections to quasi-localized modes and their power spectrum here.

In Supplementary Fig., \ref{fig:dispersion_fL_fT_N_4e3_f0_log_merge}, a heat map of both the longitudinal mode and transverse mode using the long wavelength modes of the effective dynamical matrix is given for the 2dmKA model for a system size of $N = 4000$.

\rev{\section{Non-Linear Phonon Dispersion - Vibrational Density of States on Fractals}
\begin{figure}[!h]
	\includegraphics[width=.5\textwidth]{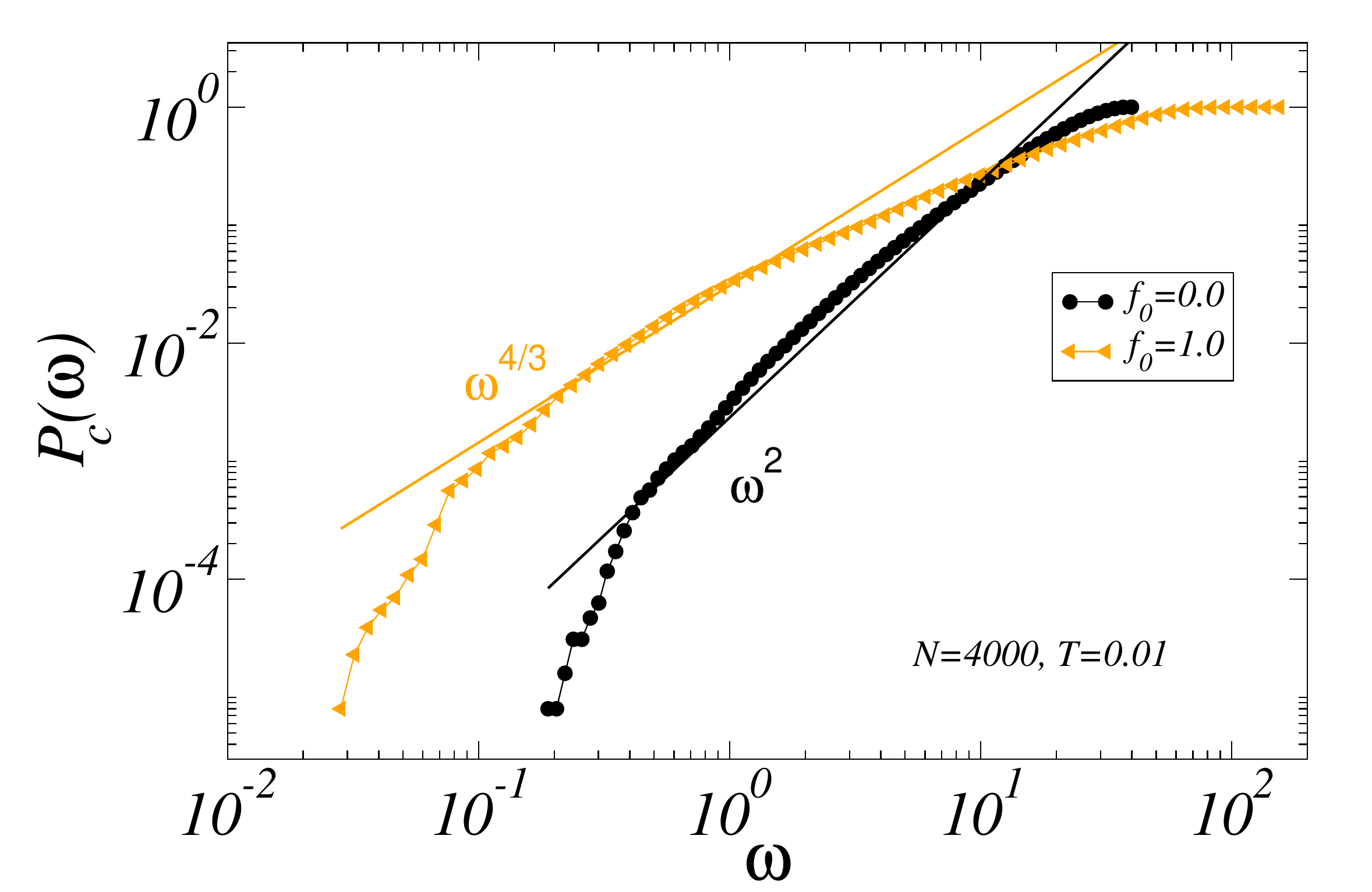}
	\caption{Cumulative distribution of vibrational density of states (VDoS) for active and passive systems are shown with the passive system showing usual Debye behaviour of $P_c(\omega) \sim \omega^2$ in 2D whereas the same system in the presence of active force shows a power-law behaviour which is much weaker than the passive system. We have put a $P_c(\omega) \sim \omega^{4/3}$ line for guide to eye. See text for detailed discussions.}
	\label{fig:VDoS_plot}
\end{figure} 
About the reason for non-linear phonon dispersion, unfortunately we don't have any microscopic theory in hand to explain this non-linear behaviour but we agree with the referee that if one can motivate a possible reason for exponent being bigger than one from a general principle would be really fascinating. One possible way non-linear phonon dispersion relation can be rationalized is to assume that the underlying structure that supports the phononic excitations in these non-equilibrium systems is a fractal in nature. Now in Ref.\cite{Orbach1982}, it was argued that density of states on a fractal can be very different compared to the usual solids and there are evidences  of VDoS going as $\sim \omega^{1/3}$ (see \cite{YakuboNakayama1987a, YakuboNakayama1989, RMP_Orbach1994} on percolating network in $d=2$). If one assumes such a VDoS and derives the phonon dispersion relation from that then one will obtain $\omega \sim q^{3/2}$ with $q$ being the wave vector. Lets assume that
\begin{equation}
\omega(q) = C_s q^{\alpha},
\label{phononDis}
\end{equation} 
where $C_s$ is the sound velocity. Now the relation between DoS and the phonon dispersion for a finite size system of linear size $L$ in 2D can be written as
\begin{equation}
\mathcal{D}(\omega) d\omega = 2 \left ( \frac{L}{2\pi}\right)^2 2\pi q dq,
\end{equation}
now if we use Eq.\ref{phononDis} and write the DoS in terms of $\omega$, then we will have
\begin{equation}
\mathcal{D}(\omega) \sim \frac{L^2}{\pi} \left ( \frac{\omega}{C_s}\right)^{\frac{2-\alpha}{\alpha}}.
\end{equation} 
If we assume that $\alpha \simeq 1.5$, then we get $\mathcal{D}(\omega) \sim \omega^{1/3}$ and cumulative VDoS going as $\omega^{4/3}$. In Supplementary Fig.\ref{fig:VDoS_plot}, we have plotted the cumulative VDoS ($P_c(\omega)$) for passive and active system in 2D. Passive system shows the well-known Debye like relation with $P_c(\omega)\sim \omega^2$ whereas active system shows much weaker dependence on $\omega$. We put a power-law relation of $\omega^{4/3}$ as a guide to eye. The power may not be very convincingly $4/3$ but it is certainly smaller than $2$. Thus it suggests that an enhanced VDoS at small frequency naturally suggests a possible non-linear phonon dispersion relation if the mechanisms via which the spectral weight of low frequency phonon modes got enhanced does not lead to localization of the modes and the modes remain phonon like extended modes. Some of these arguments are true as our effective Hessian calculation suggests that the phonon modes remain very similar in the presence of active forces. It is just that a lot more modes became de-localized as well as their frequency became smaller. Although the question of underlying fractal network controlling the vibrational properties of these active solids remain to be validated and hopefully in future we will be able to understanding the force-chain networks in the presence of active forces and their role in determining the mechanical properties of these solids. } 

\begin{figure*}[!htpb]
\centering
\includegraphics[width=0.99\textwidth]{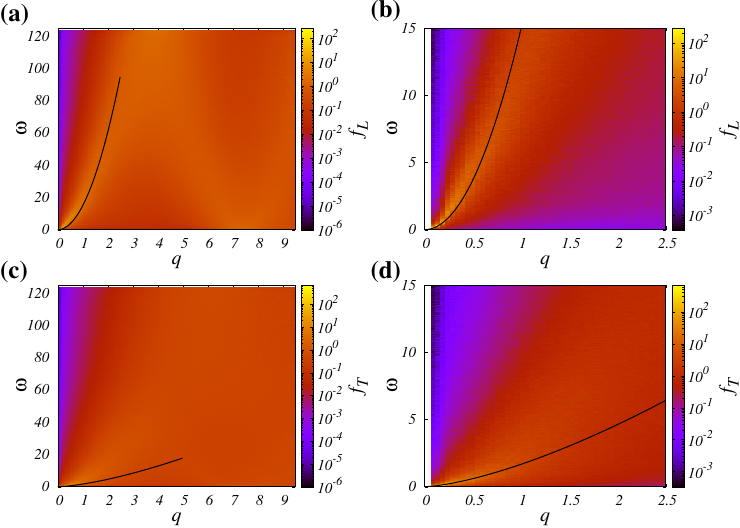}
\caption{Heat map: (a) \& (b) are dispersion relations for longitudinal spectral function ($f_L$), (c) \& (d) are dispersion relations for transverse spectral function ($f_T$). The system size N=4000 with activity $f_0=1.0$ shows deviation from the well-known linear phonon dispersion ($\omega \propto q$). For the active system, it clearly shows that the dispersion relation follows power-law behaviour ($\omega \propto q^{\alpha}$) in the small q regime, whereas for active system $\alpha > 1$ in both longitudinal and transverse spectrum. The power-law for $f_L$ is $\omega\propto q^{2.02}$ and for $f_T$ is $\omega\propto q^{1.47}$. All plots are for the 2dmKA model.}
\label{fig:dispersion_fL_fT_N_4e3_f0_log_merge}
\end{figure*}
\begin{figure*}[!htpb]
\centering
\includegraphics[width=0.98\textwidth]{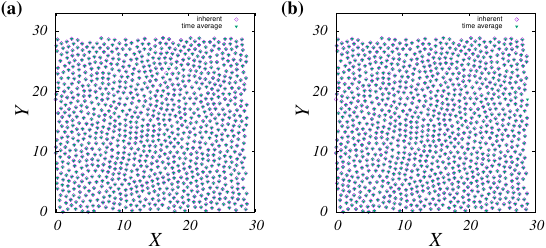}
\caption{The minimized potential energy or inherent structure (IS) state of the system and its time-averaged positions for t=5000, starting from the same minimized state. IS (circle) and time average (down triangle) are done for N=1000 system size for passive ($f_0=0.0$, (a)) and active system ($f_0=1.0$, (b)). All plots are for the 2dmKA model. }
\label{fig:inherent_t_avg_position_N_1e3_f0_2dmKA_merge}
\end{figure*}

\begin{figure*}[!htpb]
\centering
\includegraphics[width=0.92\textwidth]{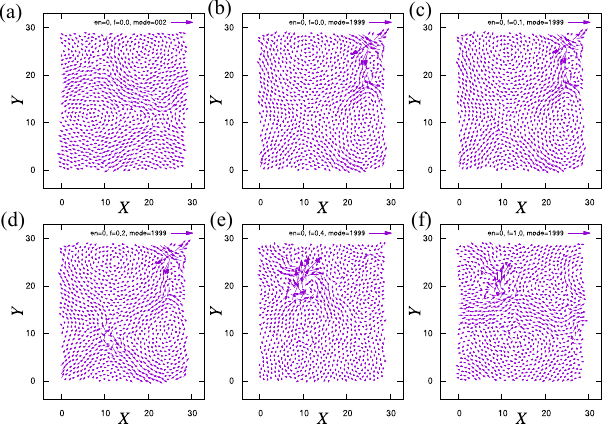}
\caption{Eigenvectors corresponding to the lowest eigenmode for activity ($f_0$) (a \& b) 0.0, (c) 0.1, (d) 0.2, (e) 0.4, (f) 1.0. Where (a) is computed using Hessian of the Inherent Structure and (b-f) is calculated using a displacement-displacement correlation matrix. All plots are for the 2dmKA model.}
\label{fig:Vect_field_N_1e3_f0_2dmKA}
\end{figure*}
\section{Inherent Structure and Eigen modes}
In Supplementary Fig. \ref{fig:inherent_t_avg_position_N_1e3_f0_2dmKA_merge}, we show that the minimized potential energy or inherent structure (IS) state of the system is close to the state achieved by the time-averaged positions over t=5000 for system size N=1000. This time-averaged position and IS state will be maintained as long as the MSD of the system of particles does not break the cage to go into the diffusive regime.

Now, starting from the same energy-minimized state of the passive system, we systematically increase the activity in the system to observe the effect on the eigenvectors. In Supplementary Fig. \ref{fig:Vect_field_N_1e3_f0_2dmKA}, we show that
different localized modes are present in the system for active as well as passive systems when we use displacement-displacement correlation matrix, but it is not present for Hessian matrix for passive system. As eigen mode changes to lower values due to activity, we can expect different hybridized eigenvectors at lower eigenmodes.

\rev{
\section{MSD calculation using Linear Response and Active Elasticity Formalisms:}
\begin{widetext}
\begin{equation}
	\zeta \delta \dot{\mathbf{r}_i} = \zeta v_0 \hat{\mathbf{n}}_i - \sum_j \mathbf{K}_{ij} \cdot \delta \mathbf{r}_j
\end{equation}
where, $\hat{\mathbf{n}}_i$ is the noise and $\mathbf{K}_{ij}$ is the Dynamical matrix.\\
N d-dimensional vectors $\left( \mathbf{\xi}_1^\nu, \dots,  \mathbf{\xi}_N^\nu \right)$, where, $\nu = 1$ to $dN$\\
So,
\begin{eqnarray}
    \delta \mathbf{r}_i &=& \sum_{\nu=1}^{dN} a_\nu \mathbf{\xi}_i^{\nu} \\
\zeta \dot{a_\nu} &=& -\lambda_\nu a_\nu + \zeta v_0 \sum_{i=1}^{dN} \hat{\mathbf{n}_i} \cdot \mathbf{\xi}_i^{\nu} = -\lambda_\nu a_\nu + \eta_\nu
\end{eqnarray}
where, $\eta_\nu$ is projection of the self propulsion force onto normal mode $\nu$ and $\sum_{i=1}^N \mathbf{\xi}_i^{\nu} \cdot \mathbf{\xi}_i^{\nu^\prime} = \delta_{\nu, \nu^\prime}$\\
Now, from the uncoupled equation
\begin{eqnarray}
    \zeta \dot{a_\nu} &=& -\lambda_\nu a_\nu + \eta_\nu \\
    \zeta \frac{d a_\nu}{dt} + \lambda_\nu a_\nu &=& \eta_\nu (t) \\
    \frac{d a_\nu}{dt} + \frac{\lambda_\nu}{\zeta} a_\nu &=& \frac{\eta_\nu (t)}{\zeta} \\
    \frac{d}{dt} \left( a_\nu e^{\frac{\lambda_\nu}{\zeta}t}\right) &=& \frac{\eta_\nu}{\zeta} e^{\frac{\lambda_\nu}{\zeta}t} \\
    a_\nu (t) e^{\frac{\lambda_\nu}{\zeta}t} &=& \int_0^t dt^\prime \frac{\eta_\nu}{\zeta} e^{\frac{\lambda_\nu}{\zeta}t^\prime} + a_\nu (0) \\
    a_\nu &=& \int_0^t dt^\prime \frac{\eta_\nu (t^\prime)}{\zeta} e^{-\frac{\lambda_\nu}{\zeta}(t-t^\prime)} + a_\nu (0) e^{-\frac{\lambda_\nu}{\zeta}t}
\end{eqnarray}
Now in the $t \rightarrow \infty$ then we get
\begin{eqnarray}
	\left< a_\nu^2 \right> 
	&=& \frac{1}{\zeta^2} \int_0^\infty dt^{\prime} \int_0^\infty dt^{\prime\prime} \left< \eta_\nu (t^{\prime}) \eta_\nu^{\prime} (t^{\prime\prime}) \right>  e^{-\frac{\lambda_\nu}{\zeta}(t-t^{\prime})} e^{-\frac{\lambda_\nu}{\zeta}(t-t^{\prime\prime})} \\
	&=& \frac{1}{\zeta^2} \int_0^\infty dt^{\prime} \int_0^\infty dt^{\prime\prime} \ C(t^{\prime} - t^{\prime\prime}) \ \delta_{\nu\nu^{\prime}} \  e^{-\frac{\lambda_\nu}{\zeta}(t-t^{\prime})} e^{-\frac{\lambda_\nu}{\zeta}(t-t^{\prime\prime})} \\
	&=&  e^{-2\frac{\lambda_\nu}{\zeta}t} \frac{1}{\zeta^2} \int_0^\infty dt^{\prime} \int_0^\infty dt^{\prime\prime} \ C(t^{\prime} - t^{\prime\prime}) \ \delta_{\nu\nu^{\prime}} \  e^{\frac{\lambda_\nu}{\zeta}(t^{\prime}-t^{\prime\prime})} \\
	&=&  e^{-2\frac{\lambda_\nu}{\zeta}t} \frac{1}{\zeta^2} \int_0^\infty dt^{\prime} \int_0^\infty dt^{\prime\prime} \ C(t^{\prime} - t^{\prime\prime}) \  e^{\frac{\lambda_\nu}{\zeta}(t^{\prime}+t^{\prime\prime})} \\
	&=& \frac{v_0^2 \zeta^2}{2} e^{-2\frac{\lambda_\nu}{\zeta}t} \frac{1}{\zeta^2} \int_0^\infty dt^{\prime} \int_0^\infty dt^{\prime\prime} \ e^{\frac{\lambda_\nu}{\zeta}(t^{\prime}+t^{\prime\prime})} e^{-\frac{|t^{\prime}-t^{\prime\prime}|}{\tau}} \\
	&=& \frac{v_0^2}{2} e^{-2\frac{\lambda_\nu}{\zeta}t} \left[ 2 \cdot \int_0^\infty dt^{\prime} \ e^{\frac{\lambda_\nu}{\zeta}t^{\prime}} \ e^{\frac{-t^{\prime}}{\tau}} \left[ \int_0^{t^{\prime}} dt^{\prime\prime} \ e^{\frac{\lambda_\nu}{\zeta}t^{\prime\prime}} \ e^{\frac{t^{\prime\prime}}{\tau}} \right] \right] \\
	&=& v_0^2 \left( \frac{1}{\frac{\lambda_{\nu}}{\zeta}+\frac{1}{\tau}} \right) \left( \frac{\zeta}{2 \lambda_{\nu}} \right) \\
	&=& \left( \frac{v_0^2 \zeta}{2 \lambda_{\nu}} \right) \left( \frac{\tau}{1 + \frac{\lambda_{\nu}}{\zeta} \tau} \right) \\
    &=& \frac{\zeta v_0^2 \tau}{2 \lambda_\nu \left( 1 + \frac{\lambda_\nu}{\zeta} \tau \right)}
\end{eqnarray}
Now, $\hat{H}(\mathbf{q}) = \left< \mathbf{r}(\mathbf{q}) \cdot \mathbf{r}^* (\mathbf{q}) \right>$, where, $\mathbf{r}(\mathbf{q}) = \frac{1}{N} \sum_{j=1}^N e^{i \mathbf{q}\cdot \mathbf{r}_j^0} \delta \mathbf{r}_j$
So,
\begin{eqnarray}
	\hat{\mathbf{H}}(\mathbf{q}) &=& \sum_{\nu, \nu^\prime} \left< a_\nu a_{\nu^\prime} \right> \mathbf{\xi}_\nu (\mathbf{q}) \cdot \mathbf{\xi}_{\nu^\prime}^* (\mathbf{q}) \\
    &=& \sum_\nu \frac{\zeta v_0^2 \tau}{2 \lambda_\nu \left( 1 + \frac{\lambda_\nu}{\zeta}\tau\right)} || \mathbf{\xi}_\nu (\mathbf{q}) ||^2
\end{eqnarray}
where, $\mathbf{\xi}_\nu(\mathbf{q}) = \frac{1}{N} \sum_{j=1}^N e^{i \mathbf{q}\cdot \mathbf{r}_j^0} \mathbf{\xi}_j^\nu$. So,
\begin{eqnarray}
	MSD &=& \frac{1}{N} \sum_{j=1}^N \left< | \delta \mathbf{r}_j |^2\right> \\
    &=& \sum_{\mathbf{q}} \hat{H} (\mathbf{q}) \\
	&=& \left(\frac{L}{2\pi}\right)^d \int d^d \mathbf{q} \sum_\nu \frac{\zeta v_0^2 \tau}{2 \lambda_\nu \left( 1 + \frac{\lambda_\nu}{\zeta} \tau \right)} || \mathbf{\xi}_\nu (\mathbf{q}) ||^2
\end{eqnarray}
	Now, $\mathbf{u}(\mathbf{r}, t) = \frac{1}{(2\pi)^(d+1)} \int d^d \mathbf{q} \int d \omega \ \tilde{\mathbf{u}}(\mathbf{q}, \omega) e^{-i \mathbf{q}\cdot \mathbf{r} -i \omega t}$ and $\tilde{\mathbf{u}}(\mathbf{q}, \omega) = \int d^d \mathbf{r} \int dt \ \mathbf{u}(\mathbf{r}, t) e^{i \mathbf{q}\cdot \mathbf{r} + i \omega t}$\\
Now from, $\zeta \dot{\mathbf{u}} = \zeta v_0 \hat{n} + \mathbf{\nabla} \cdot \hat{\sigma}$ we get, $-i \zeta \omega \tilde{\mathbf{u}}(\mathbf{q}, \omega) = \tilde{\mathbf{F}}^{act} (\mathbf{q}, \omega) - \mathbf{D} (\mathbf{q}) \tilde{\mathbf{u}}(\mathbf{q}, \omega)$, where,
\begin{equation}
	\tilde{\mathbf{F}}^{act} (\mathbf{q}, \omega) = \zeta v_0 \int_{-\infty}^{\infty} d^d \mathbf{r} \int_{-\infty}^{\infty} dt \ \hat{\mathbf{n}}(\mathbf{r}, t) \ e^{i \mathbf{q}\cdot \mathbf{r} + i \omega t}
\end{equation}
So, coorelation of active noise in continuum limit
\begin{eqnarray}
	\left< \tilde{\mathbf{F}}^{act} (\mathbf{q}, \omega) \cdot \tilde{\mathbf{F}}^{act} (\mathbf{q}^{\prime}, \omega^\prime) \right> &=& \zeta^2 v_0^2 \int_{-\infty}^{\infty} dt \int_{-\infty}^{\infty} dt^\prime \int d^d \mathbf{r} \int d^d \mathbf{r ^\prime} e^{i \omega t} e^{i \mathbf{q} \cdot \mathbf{r}} e^{i \omega^\prime t^\prime} e^{i \mathbf{q}^{\prime} \cdot \mathbf{r}^{\prime}} \left< \hat{\mathbf{\eta}}(\mathbf{r}, t) \cdot \hat{\mathbf{\eta}}(\mathbf{r}^{\prime}, t^{\prime}) \right> \\
	&=& a^d \zeta^2 v_0^2 \int_{-\infty}^{\infty} dt \int_{-\infty}^{\infty} dt^\prime \int_{-\infty}^{\infty} d^d \mathbf{r} \int_{-\infty}^{\infty} d^d \mathbf{r^\prime} e^{i \omega t} e^{i \mathbf{q} \cdot \mathbf{r}} e^{i \omega^\prime t^\prime} e^{i \mathbf{q}^{\prime} \cdot \mathbf{r}^{\prime}} \delta(\mathbf{r}-\mathbf{r}^{\prime}) e^{-\frac{|t-t^{\prime}|}{\tau}} \\
	&=& (2 \pi)^{(d+1)} a^d \zeta^2 v_0^2 \frac{2 \tau}{1 + (\tau \omega)^2} \delta(\mathbf{q}+ \mathbf{q}^{\prime})\delta (\omega + \omega^{\prime} )
\end{eqnarray}
In the above these are Dirac delta in continuum limit, to match with simulations we have to change it to Kronecker delta as
\begin{equation}
\delta (\mathbf{q} + \mathbf{q}^{\prime}) \rightarrow \frac{1}{(\Delta q)^d} \delta_{\mathbf{q}^{\prime}, -\mathbf{q}}
\end{equation}
with $\Delta q = \frac{2 \pi}{L}$. Transformation from continuum limit to finite size $\frac{1}{(2 \pi)^d} \int d^d \mathbf{q} \rightarrow \frac{1}{Na^d} \sum_{\mathbf{q}}$, $\int d^d \mathbf{r} \rightarrow a^d \sum_{\mathbf{r}}$ where $N = \frac{L^d}{a^d}$. For square lattice $\mathbf{q} = \left( \frac{2 \pi m_1}{L} , \frac{2 \pi m_2}{L}, \ldots, \frac{2 \pi m_d}{L} \right)$ where, $0 \leq m_1, m_2, \ldots , m_d \leq \frac{L}{a} -1$.\\
Relation between discrete and continuous fourier transform $\tilde{\mathbf{u}}(\mathbf{q}, t) = a^d \mathbf{u}(\mathbf{q}, t)$. So coorelation of active noise in discreat space becomes
\begin{equation}
\left< \tilde{\mathbf{F}}^{act} (\mathbf{q}, \omega) \cdot \tilde{\mathbf{F}}^{act} (\mathbf{q}^{\prime}, \omega^{\prime}) \right> = 2 \pi N \zeta^2 v_0^2 \frac{2 \tau}{1 + (\tau \omega)^2} \delta(\omega + \omega^\prime)
\end{equation}
Now, $\tilde{\mathbf{u}}(\mathbf{q}, \omega)$ can be written as $\tilde{\mathbf{u}}(\mathbf{q}, \omega) = \tilde{\mathbf{u}}_L(\mathbf{q}, \omega) \hat{q} + \tilde{\mathbf{u}}_T(\mathbf{q}, \omega) \hat{q}^\perp$.\\
So,
\begin{eqnarray}
    -i \zeta \omega \tilde{\mathbf{u}}_L (\mathbf{q}, \omega) &=& \tilde{\mathbf{F}}^{act} (\mathbf{q}, \omega) \cdot \hat{q} - (B+\mu)q^2 \tilde{\mathbf{u}}_L (\mathbf{q}, \omega) \\
    -i \zeta \omega \tilde{\mathbf{u}}_T (\mathbf{q}, \omega) &=& \tilde{\mathbf{F}}^{act}
    (\mathbf{q}, \omega) \cdot \hat{q}^\perp - \mu q^2 \tilde{\mathbf{u}}_T (\mathbf{q}, \omega)
\end{eqnarray}
where,  $\tilde{\mathbf{u}}_L (\mathbf{q}, \omega) = \frac{\tilde{\mathbf{F}}_L^{act}(\mathbf{q}, \omega)}{-i\zeta \omega +(B+\mu)q^2}$ and $\tilde{\mathbf{u}}_T (\mathbf{q}, \omega) = \frac{\tilde{\mathbf{F}}_T^{act}(\mathbf{q}, \omega)}{-i\zeta \omega +\mu q^2}$\\
Now, $\left<\tilde{\mathbf{u}}(\mathbf{q}, \omega) \cdot \tilde{\mathbf{u}}(\mathbf{q}^\prime, \omega^\prime)\right> = \left<
\tilde{\mathbf{u}}_L(\mathbf{q}, \omega) \cdot \tilde{\mathbf{u}}_L(\mathbf{q}^\prime, \omega^\prime) \right> + \left<
\tilde{\mathbf{u}}_T(\mathbf{q}, \omega) \cdot \tilde{\mathbf{u}}_T(\mathbf{q}^\prime, \omega^\prime) \right>$ and $ \left< \tilde{\mathbf{F}}_L^{act} (\mathbf{q}, \omega) \cdot \tilde{\mathbf{F}}_L^{act} (\mathbf{q^\prime}, \omega^\prime) \right> = \left< \tilde{\mathbf{F}}_T^{act} (\mathbf{q}, \omega) \cdot \tilde{\mathbf{F}}_T^{act} (\mathbf{q^\prime}, \omega^\prime) \right> = \frac{1}{2} \left< \tilde{\mathbf{F}}^{act} (\mathbf{q}, \omega) \cdot \tilde{\mathbf{F}}^{act} (\mathbf{q^\prime}, \omega^\prime) \right>$\\
So,
\begin{eqnarray}
    \left< \tilde{\mathbf{u}}_L(\mathbf{q}, \omega) \cdot \tilde{\mathbf{u}}_L(\mathbf{q}^\prime, \omega^\prime) \right> &=& \frac{\left< \tilde{\mathbf{F}}_L^{act} (\mathbf{q}, \omega) \cdot \tilde{\mathbf{F}}_L^{act} (\mathbf{q^\prime}, \omega^\prime) \right>}{\zeta^2 \omega^2 + (B+\mu)^2 q^4} \\
    &=& \frac{1}{2} \frac{\left< \tilde{\mathbf{F}}^{act} (\mathbf{q}, \omega) \cdot \tilde{\mathbf{F}}^{act} (\mathbf{q^\prime}, \omega^\prime) \right>}{\zeta^2 \omega^2 + (B+\mu)^2 q^4} \\
	&=& \frac{(2\pi)^{(d+1)} a^d \zeta^2 v_0^2 \tau}{\left[ \zeta^2 \omega^2 + (B+\mu)^2 q^4 \right]\left[
    1+(\tau \omega)^2 \right]} \delta(\mathbf{q}+\mathbf{q}^{\prime}) \delta(\omega + \omega^{\prime})
\end{eqnarray}
Similarly,
\begin{equation}
	\left< \tilde{\mathbf{u}}_T(\mathbf{q}, \omega) \cdot \tilde{\mathbf{u}}_T(\mathbf{q}^{\prime}, \omega^\prime) \right> = \frac{(2\pi)^{(d+1)} a^d \zeta^2 v_0^2 \tau}{\left[ \zeta^2 \omega^2 + \mu^2 q^4 \right]\left[
1+(\tau \omega)^2 \right]} \delta(\mathbf{q}+\mathbf{q}^{\prime}) \delta(\omega + \omega^{\prime})
\end{equation}
Now,
\begin{eqnarray}
\left< \tilde{\mathbf{u}}_L(\mathbf{q}, t) \cdot \tilde{\mathbf{u}}_L(\mathbf{q}^{\prime}, t) \right> &=& \frac{1}{(2\pi)^d}\int_{-\infty}^\infty d \omega \int_{-\infty}^\infty d \omega^\prime e^{-i(\omega +\omega^\prime)t} \left< \tilde{\mathbf{u}}_L(\mathbf{q}, \omega) \cdot \tilde{\mathbf{u}}_L(\mathbf{q}^{\prime}, \omega^{\prime}) \right> \\
	&=& (2 \pi)^{(d-1)} a^d \zeta^2 v_0^2 \tau \delta(\mathbf{q}+\mathbf{q}^{\prime}) \int_{-\infty}^{\infty} d \omega \frac{1}{\left[ (B+\mu)^2 q^4 +\zeta^2 \omega^2 \right]\left[ 1+(\tau \omega)^2 \right]}\\
	&=& (2 \pi)^{(d-1)} a^d \zeta^2 v_0^2 \tau \delta(\mathbf{q}+\mathbf{q}^{\prime}) \left[\frac{\pi}{q^2 \left( (B+\mu)\zeta +(B+\mu)^2 \tau q^2 \right)} \right] \\
	&=& \frac{(2 \pi)^d a^d \zeta^2 v_0^2 \tau \delta(\mathbf{q}+\mathbf{q}^{\prime})}{2} \left[\frac{1}{q^2 \left( (B+\mu)\zeta +(B+\mu)^2 \tau q^2 \right)} \right]
\end{eqnarray}
and,
\begin{equation}
	\left< \tilde{\mathbf{u}}_T(\mathbf{q}, t) \cdot \tilde{\mathbf{u}}_T(\mathbf{q}^{\prime}, t) \right> =  \frac{(2 \pi)^d a^d \zeta^2 v_0^2 \tau \delta(\mathbf{q}+\mathbf{q}^{\prime})}{2} \left[\frac{1}{q^2 \left( \mu\zeta + \mu^2 \tau q^2 \right)} \right]
\end{equation}
So,
\begin{equation}
	\left< \tilde{\mathbf{u}}(\mathbf{q}, t) \cdot \tilde{\mathbf{u}}(\mathbf{q}^{\prime}, t) \right> = \frac{(2 \pi)^d a^d \zeta^2 v_0^2 \tau}{2 q^2}\left[ \frac{1}{(B+\mu)\zeta +(B+\mu)^2 \tau q^2} + \frac{1}{\mu \zeta + \mu^2 \tau q^2}\right] \delta (\mathbf{q} + \mathbf{q}^{\prime})
\end{equation}
Now,
\begin{eqnarray}
	\left<\mathbf{u}(\mathbf{r},t) \cdot \mathbf{u}(\mathbf{r},t)\right> 
	&=& \frac{1}{(2 \pi)^{2d}} \int d^d \mathbf{q} \int d^d \mathbf{q}^{\prime} \left< \tilde{\mathbf{u}}(\mathbf{q}, t) \cdot \tilde{\mathbf{u}}(\mathbf{q}^\prime, t) \right> e^{-i (\mathbf{q}+\mathbf{q}^{\prime}) \cdot \mathbf{r}} \\
	&=& \frac{a^d \zeta^2 v_0^2 \tau}{2 (2 \pi)^d q^2} \int d^d \mathbf{q} \int d^d \mathbf{q^{\prime}} \left[ \frac{1}{(B+\mu)\zeta +(B+\mu)^2 \tau q^2} + \frac{1}{\mu \zeta + \mu^2 \tau q^2}\right] \delta (\mathbf{q} + \mathbf{q}^{\prime}) e^{-i (\mathbf{q}+\mathbf{q}^{\prime}) \cdot \mathbf{r}} \\
    &=& \frac{a^d \zeta^2 v_0^2 \tau}{2 (2 \pi)^d q^2} \int \frac{d^d \mathbf{q}}{q^2} \left[ \frac{1}{(B+\mu)\zeta + (B+\mu)^2 \tau q^2} + \frac{1}{\mu \zeta + \mu^2 \tau q^2} \right]
\end{eqnarray}
Now for the finite $\tau$ limit,
\begin{eqnarray}
\left<|\mathbf{u}(\mathbf{r},t)|^2\right> 
	&=& \left(\frac{1}{2\pi}\right)^{d/2} a^{d} \frac{(v_0 \tau)^2}{2} \frac{1}{\Gamma(d/2)} \int dq \frac{q^{d-1}}{q^2}  \left[\frac{1}{\xi_L^2(1+\xi_L^2 q^2)}
+ \frac{1}{\xi_T^2(1+\xi_T^2 q^2)} \right]
\end{eqnarray}
For 2d system the mean square displacement at finite $\tau$ is given by,
\begin{eqnarray}
\left<|\mathbf{u}(\mathbf{r},t)|^2\right>
	&=& \frac{1}{4\pi} a^2 v_0^2 \zeta \tau \cdot \frac{1}{2(B+\mu)} \left[ 2 \ln(L/a) - \ln \left(1+\left(\frac{2\pi}{L} \xi_L\right)^2 \right) + \ln \left(1+\left(\frac{2\pi}{a} \xi_L\right)^2 \right) \right] \notag \\
	&+& \frac{1}{4\pi} a^2 v_0^2 \zeta \tau \cdot \frac{1}{2\mu} \left[ 2 \ln(L/a) - \ln \left(1+\left(\frac{2\pi}{L} \xi_T\right)^2 \right) + \ln \left(1+\left(\frac{2\pi}{a} \xi_T\right)^2 \right) \right].
\end{eqnarray}
If we take $L \rightarrow \infty$ of the above equation, then we get 
\begin{align}
	\left<|\mathbf{u}(\mathbf{r},t)|^2\right> = 
	\frac{1}{4\pi} a^2 v_0^2 \zeta \tau \cdot \left[ \frac{1}{(B+\mu)} + \frac{1}{\mu}\right] \ln(L/a)
\end{align}
\begin{figure*}[!htb]
\centering
        \includegraphics[scale = 1.2]{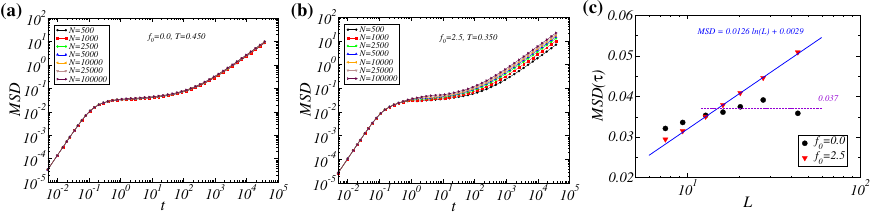}
        \caption{{\bf Mermin-Wagner-Hohenberg fluctuation in 3D:} (a) Mean Square Displacement (MSD) for various system sizes for passive 3dKA model. (b) The same plot with activity $f_0 = 2.5$. Notice the change in the MSD plateau with increasing system size for active cases. (c) shows the MSD plateau as a function of the logarithm of system size ($L$). $MSD(\tau) \sim \log{(L)}$ for $f_0 = 2.5$ and MSD plateau tends to saturate for the larger system size for a passive system.}
        \label{3dResults}
\end{figure*}
\end{widetext}
}

\begin{figure*}[!htb]
\centering
        \includegraphics[scale = 1.2]{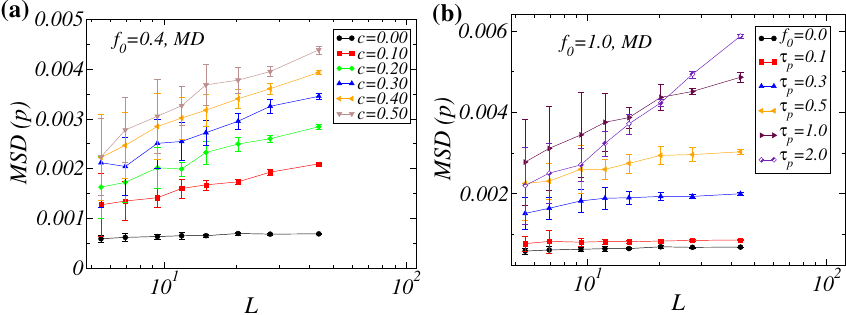}
	\caption{{\bf Mermin-Wagner-Hohenberg fluctuation in 3D by changing c \& $\tau_p$:} (a) Mean squared displacement (MSD) plateau diverges logarithmically for higher activity, and for the passive case it does not shows any divergence. We changed the concentration of active particles by keeping constant per particle force $f_0 = 0.4$ and $\tau_p = 1.0$. (b) Mean squared displacement (MSD) plateau again diverges logarithmic with changing $\tau_p$, and for the passive case and lower $\tau_p$ values it does not shows any divergence at large system size limit. The activity is changed by changing persistent time $\tau_p$ by keeping per particle force $f_0 =1.0$ and active particle concentration $c=0.1$ constant. Error bars in the figure panels are measured by computing the standard deviation (SD) of fluctuations in various statistically independent simulations.}
        \label{fig:msdvsL_C_taup_3dKA_merge}
\end{figure*}

\rev{
\section{MSD plateau for 3D system in supercooled regime:}
Here we show the MWH effects in the 3D at supercooled temperature regime. We plotted MSD value at the plateau for various systems sizes ranging from $500$ to $100000$ for activity $f_0 = 2.5$ and compared that with the passive case as shown in Supplementary Fig. \ref{3dResults}(A) and (B). Notice the drastic increase of plateau value of MSD with increasing system size for $f_0 = 2.5$ case as compared to the passive system where MSD plateau saturates beyond a system size very clearly as shown in Supplementary Fig. \ref{3dResults}(C). Clear logarithmic divergence of MSD plateau with increasing system size for active glasses is in remarkable agreement with the predictions from effective medium theory. In addition we have also studied the system size dependence of MSD plateau for different activity for 3d system by changing concentration and persistence time as shown in Supplementary Fig. \ref{fig:msdvsL_C_taup_3dKA_merge}(a) and (b) respectively at a constant per particle force ($f_0$ =0.4), and for changing persistent time shown in \ref{fig:msdvsL_C_taup_3dKA_merge}(b), with constant per particle force and active particle concentration fixed at $f_0 =1.0$, $c=0.1$. For both the conditions, the mean squared displacement (MSD) plateau diverges logarithmically with system size.}

\bibliographystyle{apsrev4-2}
\bibliography{MerminWagnerSupplementary}